\documentclass[
reprint,
groupedaddress,
showpacs,
showkeys,
amsmath, amssymb,
aps,
prl,
]{revtex4-2}

\usepackage{xcolor}
\usepackage{graphicx}
\usepackage{dcolumn}
\usepackage{bm}
\usepackage[colorlinks=true, allcolors=blue]{hyperref}
\usepackage{comment}
\usepackage[normalem]{ulem}
\graphicspath{{figures/}}
\usepackage[separate-uncertainty=false]{siunitx}
\usepackage{braket}
\usepackage{xspace}
\usepackage{xstring}

\newcommand{\mtwo}{$\ket{2, -2}$\xspace}
\newcommand{\mone}{$\ket{1,  -1}$\xspace}
\newcommand{\two}{$\up$\xspace}

\newcommand{\down}{\ensuremath{\ket{\downarrow}}\xspace}
\newcommand{\up}{\ensuremath{\ket{\uparrow}}\xspace}

\newcommand{\suppress}[1]{\textcolor{brown}{}}

\newcommand*{\aref}[1]{%
	\IfBeginWith{#1}{eq:}{Eq.~\eqref{#1}}{}
	\IfBeginWith{#1}{fig:}{Fig.~\ref{#1}}{}%
	\IfBeginWith{#1}{tab:}{Table~\ref{#1}}{}%
	\IfBeginWith{#1}{appendix:}{Appendix~\ref{#1}}{}%
	\IfBeginWith{#1}{sec:}{Section~\ref{#1}}{}%
	}

\usepackage{tikz}
\newcommand*\circled[1]{\tikz[baseline=(char.base)]{
    \node[shape=circle, draw, inner sep=1pt, 
        minimum height=12pt] (char) {#1};}}	
	
\begin{document}

\title{Observation of Temperature Effects on \\ False Vacuum Decay in Atomic Quantum Gases}

\author{Riccardo Cominotti$^1$}
\thanks{These two authors contributed equally to this work.}
\author{Cosetta Baroni$^{1,2}$}
\thanks{These two authors contributed equally to this work.}
\author{Chiara Rogora$^1$}
\author{Diego Andreoni$^1$}
\author{Giacomo Guarda$^1$}
\author{Giacomo Lamporesi$^1$}
\email{giacomo.lamporesi@ino.cnr.it}
\author{Gabriele Ferrari$^1$}
\author{Alessandro Zenesini$^1$}

\affiliation{$^1$ Pitaevskii BEC Center, CNR-INO and Dipartimento di Fisica, Universit\`a di Trento, 38123 Trento, Italy, and Trento Institute for Fundamental Physics and Applications, INFN, 38123 Trento, Italy\\
$^2$ Institute for Quantum Optics and Quantum Information (IQOQI) and Institute for Experimental Physics, University of Innsbruck 6020 Innsbruck, Austria}

\begin{abstract}
Temperature plays a crucial role in metastable phenomena, not only by contributing to determine the state (phase) of a system, but also ruling the decay probability to more stable states. Such a situation is
encountered in many different physical systems, ranging from chemical reactions to magnetic structures. The characteristic decay timescale is not always straightforward to estimate since it depends on the microscopic details of the system.
A paradigmatic example in quantum field theories is the decay of the false vacuum, manifested via the nucleation of bubbles.
In this paper, we measure the temperature dependence of the timescale for the false vacuum decay mechanism in an ultracold atomic quantum spin mixture which exhibits ferromagnetic properties. Our results show that the false vacuum decay rate scales with temperature as predicted by the finite-temperature extension of the instanton theory, and confirm atomic systems as an ideal platform where to study out-of-equilibrium field theories.

\end{abstract}

\maketitle

Bubble nucleation is a macroscopic phenomenon associated to non-equilibrium dynamics across a phase transition.
Despite its importance in several fields of research, from physics \cite{Franks1982, Devoto2022} to biology \cite{devries1969}, from engineering \cite{herlach2001, Gallo2021} to meteorology \cite{Kulmala2013}, it still remains poorly understood. Bubbles form inside an initially metastable system as a way to reach a lower energy minimum. This is well known to happen in presence of first-order phase transitions where two minima with different energy exist. However, it remains challenging to understand how the microscopic properties and the interfaces between different phases govern the stochastic bubble nucleation.

In classical systems, thermal fluctuations are responsible for random local inhomogeneities which help triggering the formation of bubbles. Fluctuations increase with temperature, but temperature can also be the variable discriminating between two possible stable phases, as it can favor either one of the two minima of the free energy of a system.
Quantum systems at zero temperature, instead, can be driven across a first-order phase transition by an external control parameter, but also by pure quantum fluctuations \cite{Sachdev2011}.

On top of this, a description of the system in terms of field theories allows to treat the quantum many-body system as a unique entity, greatly reducing the complexity of the problem, but still capturing the fundamental physical properties.
The metastability of the quantum field results from the competition between the local gain in potential energy and the cost in kinetic energy for the formation of the bubble surface, which turns into a sharp spatial variation of the field.

In such a system, the decay toward the low energy state --- the \textit{true vacuum} (TV) --- is a pure quantum process mediated by the formation of a resonant bubble, i.e., a heterogeneous field configuration having the same energy of the starting uniform metastable state. The instanton solution by Euclidean path integral method allows to calculate the decay rate of the metastable field state --- the \textit{false vacuum} (FV) --- \cite{Kobzarev1974, Coleman1977, Coleman1977erratum, Callan1977, Devoto2022} and to find universal behaviors from atomic to cosmological scales \cite{LANGER1969, Quiros1999, Fialko2015, Braden2018, Billam2019, Billam2020, Billam2021, Jenkins2024, Mazumdar2019, Hindmarsh2021}. 

\begin{figure}[b!]
    \centering
    \includegraphics[width =  \columnwidth]{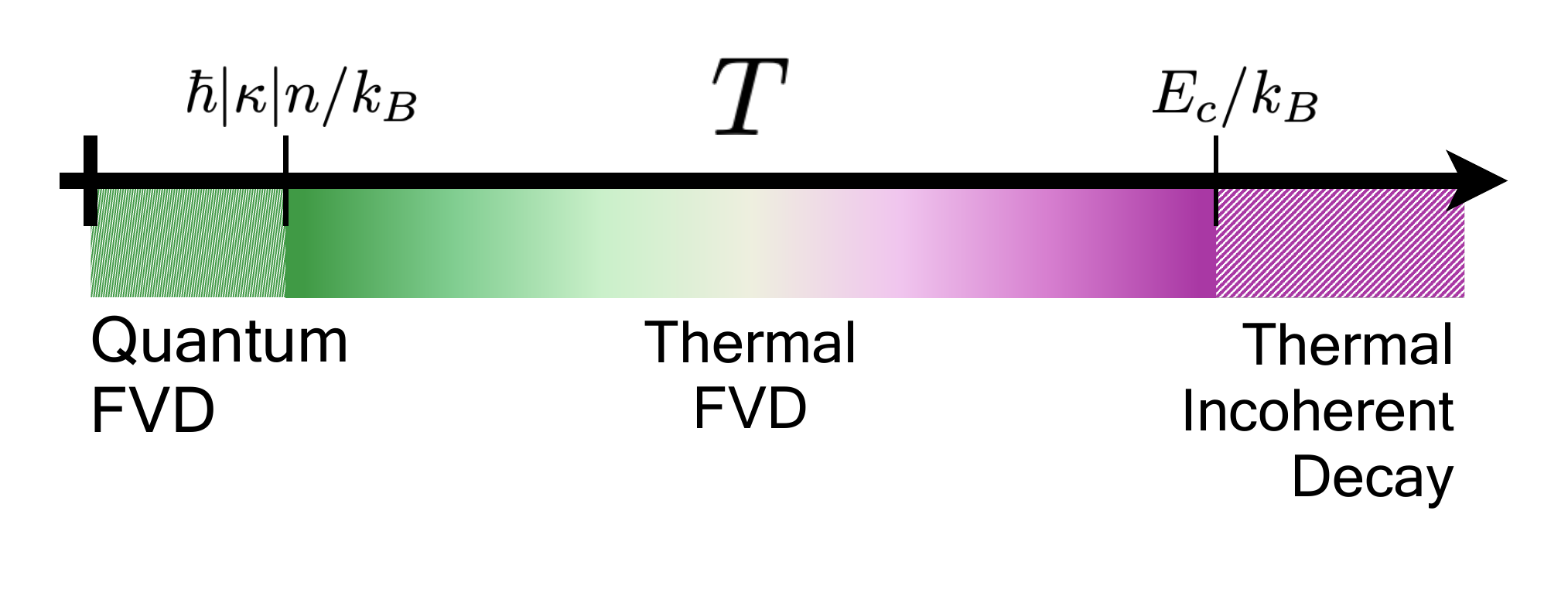}
    \caption{Relevant temperature scales in the FVD. 
    For temperatures smaller than the one associated with interactions, which in our atomic case are $\hbar|\kappa|n/k_B$ (see later), quantum fluctuations dominate in the FVD process. If the system temperature is higher but still smaller than $E_c/k_B$, the FVD is driven by thermal fluctuations. For higher temperatures, the decay is an incoherent process.}
    \label{fig:fig1}
\end{figure}

The extension of \textit{false vacuum decay} (FVD) to the thermal regime is a challenging aspect \cite{Lifshitz72, Gorokhov97} due to the non-perturbative nature of the process. Linde showed how to extend the Euclidean path integral method to finite-temperature field theories \cite{Linde1983} where thermal fluctuations are supposed to support the quantum transition to the resonant bubble. 
Defining the instanton energy as $E_c = \int [(\nabla Z)^2/2+V(Z)]\,dx$, which is a function of the spatial variation of the field order parameter $Z$ and of the field potential $V(Z)$, the characteristic decay time $\tau$ for thermally-induced false vacuum decay (FVD) in one spatial dimension is 

\begin{equation}
    \tau = A \frac{e^{\beta E_c}}{\sqrt{\beta E_c}},
\label{eq:eq1}
\end{equation}
where $A$ is a system-dependent constant and $\beta=1/k_B T$, with $k_B$ the Boltzmann constant and $T$ the system temperature. Within Linde's approach, Eq.\,\ref{eq:eq1} is valid as long as thermal fluctuations are significantly larger than quantum fluctuations, but still smaller than the instanton energy; see \aref{fig:fig1}. At very large temperatures, above $E_c/k_B$, bubble formation appears as a pure thermal process that breaks up the coherence of the field and poorly depends on the control parameter.

Experimental evidence of FVD was shown in zero-temperature spin chains with a quantum annealer \cite{Vodeb2025}, in atomic systems in optical lattices \cite{Zhu2024}, in trapped ions systems \cite{Luo2025}, and in a finite-temperature atomic system \cite{Zenesini2024} by observing the exponential character of Eq.\,\ref{eq:eq1} through variation of $E_c$. In Ref.~\cite{Zenesini2024}, the decay was attributed to thermal fluctuations rather than quantum, from a purely energetic consideration.
However, the validity of Linde's approach and the temperature dependence in the mechanism of FVD remain experimentally untested \cite{Ramos1993, Gutierrez2023}. 

\begin{figure}[b!]
    \centering
    \includegraphics[width = \columnwidth]{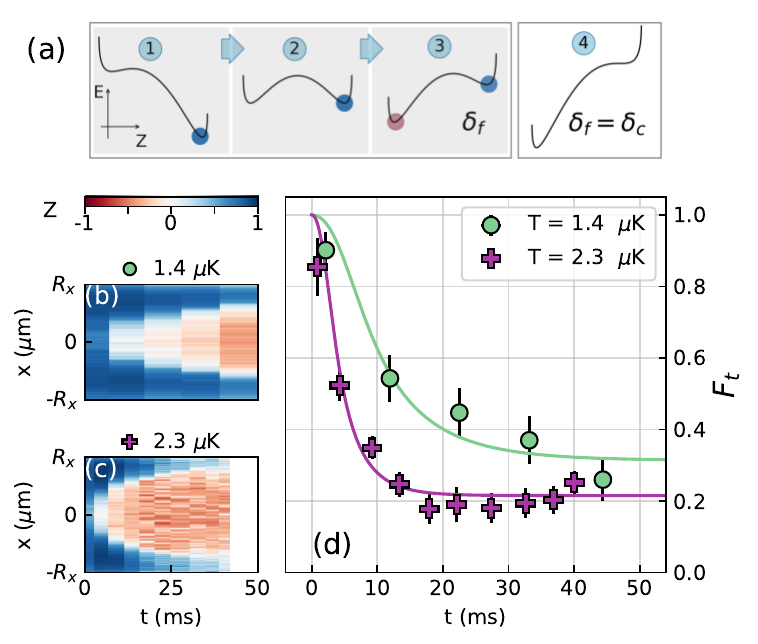}
    \caption{(a) Potential profiles during the main steps of the protocol. $\textcircled{1}$ Initial preparation in \up; $\textcircled{2}$ Symmetric configuration for $\delta = 0$; $\textcircled{3}$ Metastable state and absolute ground state at $\delta_f$. $\textcircled{4}$ Potential profile at the critical detuning $\delta_c$. (b),(c) Evolution of the system average magnetization $\langle Z(x)\rangle$ for increasing waiting time $t$. 
    Two different temperature regimes are shown for similar $(\delta_f-\delta_c)/2\pi \sim 60$ Hz, with $T_\text{exp} =$ 1.4(1)~$\mu$K (a) and 2.3(1)~$\mu$K (b). (d) Survival probability of the FV $F(t)$ in the corresponding two datasets. Error bars represent the standard error of the average magnetization.}
    \label{fig:fig2} 
\end{figure}

In this work, we test the validity of Linde's extension by observing the temperature dependence of the FVD rate in a harmonically-trapped quantum mixture \cite{Baroni2024} of ultracold sodium atoms. The atoms are allowed to populate only two internal spin states, $\ket{F,m_F}=\ket{1,-1} \equiv \down$ and $\ket{2,-2} \equiv \up$ \cite{Cominotti2023}, which are coupled by a microwave radiation with strength $\Omega$ \cite{Recati2022}.   
The mixture behavior is well described by the total density $n=(n_{\downarrow} + n_{\uparrow})$ and by the relative population imbalance in the two spin states, $Z = (n_{\uparrow} - n_{\downarrow})/ n$, where 
$n_i$ is the density of atoms in the state $\ket{i}$.
By trapping the sample in a quasi one-dimensional geometry, the dynamics of $Z$ occurs only along the direction $x$, and the three-dimensional density $n$ can be reduced to a Thomas-Fermi profile $n(x) = 2n_0(1-x^2/R_x^2)/3$ \cite{SupMat}, where $n_0$ is the peak density and $R_x$ the Thomas-Fermi radius along the axial direction.
The temperature-independent mean-field potential associated to $Z$ is \cite{Abad2013, Cominotti2023}

\begin{equation}
V(Z)= -\hbar \left( |\kappa| n Z^2 + 2 \Omega\sqrt{1-Z^2} + 2 \delta  Z \right) ,
\label{eq:eqVZ}
\end{equation}

where $\kappa \equiv (g_{\downarrow\downarrow} + g_{\uparrow\uparrow} - 2g_{\downarrow\uparrow})/2\hbar$, with $g_{ij}$ the interaction constants between two atoms in the state $\ket{i}$ and $\ket{j}$, and $\delta$ is the detuning of the radiation, keeping into account density-dependent collisional shifts $n(g{_{\downarrow\downarrow}} - g_{\uparrow\uparrow})/2\hbar$.
The system characterized by the mean-field potential of \aref{eq:eqVZ} exhibits a ferromagnetic ground state when $\kappa n/\Omega<-1$ \cite{Cominotti2023}, described by the order parameter $Z$. At $\delta = 0$, the corresponding energy profile is a symmetric double well, whereas for nonzero $|\delta|$, but still smaller than a critical value $|\delta_c|$, the degeneracy between the two minima is broken. 
By varying $\delta$ from positive to negative values, the absolute minimum of $V$ changes from positive to negative $Z$, see \circled{1}, \circled{2}, and \circled{3} in \aref{fig:fig2}(a).
This leads to a metastable state, which can decay to the ground state via the formation of \down domains (bubbles) in a bulk of \up, or vice versa. Finally, the presence of thermal fluctuations is ensured by the finite temperature of the non-condensed fraction of the atomic sample, which acts as a thermal bath. 

While in Ref.~\cite{Zenesini2024} the scaling of $\tau$ with $\delta$ and $\Omega$ was measured at a fixed temperature, in this work we use the same protocol, but act on both terms of Eq.~\ref{eq:eq1}, $\beta$ and $E_c$, by exploring a wide range of values for the initial system temperature and for different $(\delta-\delta_c)$, at a fixed value of $\Omega$. 
This procedure helps decoupling the contributions to $\tau$ arising from $\beta$ or from $E_c$.

The atoms are trapped and cooled in an elongated optical dipole trap until the gas, polarized in the $\down$ state, is partially condensed and in thermal equilibrium at a variable temperature $T_\mathrm{exp}$, depending on the final intensity of the optical trap at the end of the evaporation. To ensure identical trapping parameters for measurements at different temperatures, after a thermalization time of 500\,ms, the intensity of the optical trap is ramped up in 500\,ms to a fixed value corresponding to trapping frequencies $\{\omega_x,\omega_r\}/2\pi= \{15(1), 2020(20)\}$\,Hz in the longitudinal and transverse directions, respectively. 
Additionally, the chosen intensity corresponds to a trap depth that allows the system to be considered quasi one-dimensional for what concerns spin excitations \cite{Farolfi2021, Cominotti2023}. 
At the end of this initial preparation procedure, the sample contains about $10^6$ atoms in the hyperfine state $\down$ at temperatures ranging from 1 to 2.5 $\mu$K. This temperature lays in the range of validity of \aref{eq:eq1} \cite{Linde1983}: it is higher than the one associated to quantum fluctuations $\hbar |\kappa| n /k_B \approx$ 50\,nK and one order of magnitude lower than $E_c/k_B$, see \aref{fig:fig1}. The latter is of the order of a few tens of $\mu$K, and is dominated by the energy cost for introducing two domain walls delimiting the bubble.

Once the condensate is formed at a given temperature, we prepare the system in the metastable ferromagnetic phase \cite{SupMat}. In particular, all atoms are first transferred to the state \two by a fast microwave $\pi$-pulse without significant spurious heating.
By increasing the Zeeman splitting between the two states, we tune the detuning $\delta$ to a large positive value, and subsequently we turn on the coupling radiation at $\Omega = 2\pi \times 300$\,Hz $< |\kappa|n$, such that the prepared state corresponds to the ground state of the system. Afterwards, we linearly ramp the detuning to a negative variable value $\delta_f$  $(0>\delta_f>\delta_c)$, where the prepared state becomes metastable and the FVD occurs. For each value of $\delta_f$, we let the system evolve for a variable waiting time $t$, after which we perform independent imaging of the two atomic states. The experiment is then repeated for different $T_\mathrm{exp}$.  The adiabatic preparation ramp implies that the density and magnetization fluctuations are thermalized, suggesting that the temperature $T_\mathrm{exp}$ can be associated to both. Thermalization is also confirmed by the experimental observation that $T_\mathrm{exp}$ does not change during the waiting time and as long the state remains in the false vacuum [see Fig.~3(g) in Ref.~\cite{SupMat}].

From the pictures of each atomic sample, we evaluate the local total density $n(x)$ and the local magnetization $Z(x)$ of the condensate~\cite{SupMat}.  The measured density is used to evaluate, for each single experimental realization, the system temperature $T_\mathrm{exp}$ from the thermal tails and the parameter $|\kappa| n \approx 2\pi \times\SI{1}{kHz}$ from the condensate, which is also used to determine $\delta_c$ \cite{SupMat}.

Since the FVD manifests as the stochastic formation of bubbles of \down in \up, the time evolution of the magnetization, averaged over several realizations for given values of $T_\mathrm{exp}$ and $\delta_f$, provides information on the characteristic FVD time $\tau$.
We extract $\tau$, by first evaluating the false vacuum survival probability $F_t=(\langle Z \rangle_t -Z_{TV})/(\langle Z\rangle_0-Z_{TV})$ \cite{Lagnese2021}. Here $\langle Z \rangle_t$ stands for a spatial average of $Z(x)$ in the central $\SI{60}{\mu m}$ over several realizations at time $t$, and $Z_{TV}$ the magnetization of the true vacuum state. The experimentally extracted $F_t$  \cite{SupMat} is then fitted with the empirical function $(1-\epsilon)/\sqrt{1+(e^{t/\tau}-1)^2}+\epsilon$, which has shown to well approximate both experimental data and the numerical simulations \cite{Zenesini2024}. The parameter $\epsilon$ is introduced to take into account the minimum magnetization observable in the experiment.

In \aref{fig:fig2}(b-c), we report two exemplary data sets showing the evolution of the magnetization profiles after different waiting times $t$ and the extracted value of $F_t$ [\aref{fig:fig2}(d)]. The two datasets have different values of $T_\text{exp}$ and similar values of the control parameter $(\delta_f - \delta_c) /2\pi \sim 60$~Hz. In addition, \aref{fig:fig2}(d) highlights our procedure for extracting the decay time $\tau$ from a fit of $F_t$ with the aforementioned empirical formula.
We can see that a higher temperature favors the early formation of bubbles of $\down$ [red in the magnetization profiles of panels (b) and (c)], corresponding to a shorter decay time.

\begin{figure}[b]
    \centering
    \includegraphics[width = \columnwidth]{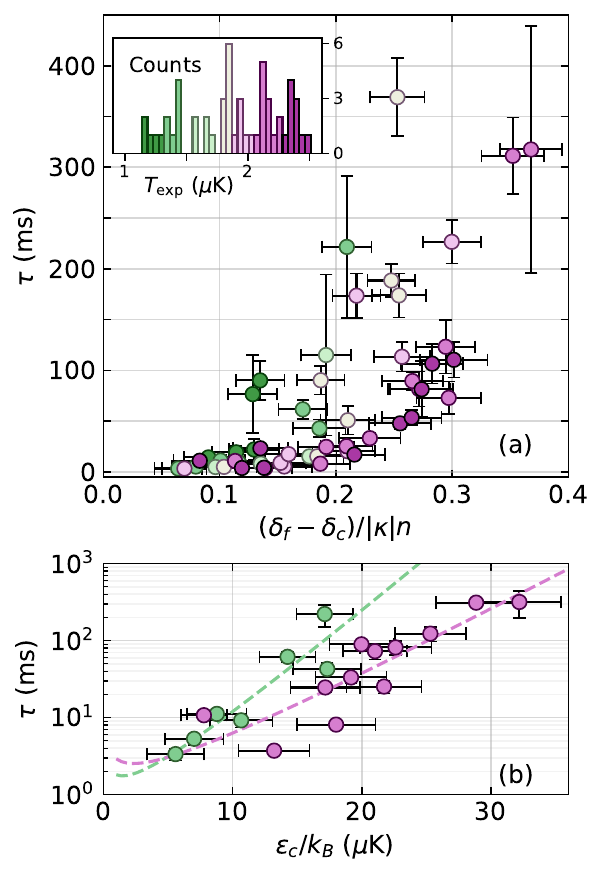}
        \caption{(a) Measured decay time $\tau$ as a function of the parameter $(\delta_f-\delta_c)/|\kappa|n$. The color scale accounts for the different temperatures of each cluster. Each point was obtained from about one hundred experimental realizations. Error bars are given by the experimental uncertainties on $(\delta_f-\delta_c)$ and $|\kappa|n$ and fit uncertainty on $\tau$. The inset shows the seven clusters in which the data are grouped according to the temperature. (b) The lin-log plot of $\tau$ as a function of $\varepsilon_c$ shows a quasi-linear trend with smaller slope for higher temperatures. The temperature for the green and purple datasets are 1.40(3) and 2.26(6) $\mu$K, respectively. Dashed lines are the  fits according to Eq.\,\ref{eq:tau-fit}.}
    \label{fig:fig3}
\end{figure}

By applying the same analysis procedure to all the collected measurements, we obtain $\tau$ for different values of the control parameter $(\delta_f - \delta_c) $, 
and of $T_\text{exp}$. These are reported in Fig.\,\ref{fig:fig3}(a) as a function of the dimensionless quantity $(\delta_f - \delta_c) /|\kappa|n$. The results have been clustered in seven different groups corresponding to different intervals of $T_\mathrm{exp}$,
as shown by the histogram in the inset of \aref{fig:fig3}(a) with a color scale going from dark green (coldest samples) to dark purple (hottest samples). 
These results show that, for any temperature, $\tau$ rapidly increases with increasing $(\delta_f-\delta_c)/|\kappa|n$, as the unbalance between the two vacua reduces. Additionally, $\tau$ for the cold samples raises faster than for the hot ones.
This behavior qualitatively confirms the expectation that a higher temperature, which is associated with stronger thermal fluctuations, boosts the FVD mechanism.

To isolate the temperature dependence in Eq.~\ref{eq:eq1}, we monitor all the experimental parameters present in the estimation of the instanton energy $E_c$ as developed in Ref.~\cite{Zenesini2024}:

\begin{equation}
    E_c=(\hbar |\kappa| n)(n_{1\text{D}} \xi_s)\left(\frac{\delta_f-\delta_c}{|\kappa| n} \right)^{\frac{5}{4}}\left(\frac{\delta_c}{|\kappa| n}\right)^{-\frac14}\mathcal{G},
    \label{eq:Ec}
\end{equation}

with $n_{1\text{D}} \approx 5\times 10^9$~atoms/cm being the one-dimensional density of the sample in the center of the trap, $\xi_s \equiv \sqrt{\hbar / 2 m |\kappa| n}\approx 0.5\,\mu \text{m}$ the spin healing length, $\mathcal{G} \equiv \mathcal{G}(\Omega/|\kappa| n)$ a dimensionless function, and $m$ the atomic mass. 
Since the dependence of $\tau$ on $\Omega/\kappa n$ for this system was not found \cite{Zenesini2024} to be consistent among simulation, experiment, and instanton theory, the expression of $\mathcal{G}$ is not precisely known. However, for this experiment, $\Omega$ is fixed and $|\kappa|n$ does not vary systematically with $T_\mathrm{exp}$, therefore we factor out the dependence of $\tau$ on $\mathcal{G}$ and consider the quantity $\varepsilon_c=E_c/\mathcal{G}$, which can be obtained from our experimental parameters. The variation of $|\kappa| n$ is instead kept into account in the direct evaluation of $\delta_c$ \cite{SupMat}, as well as in the calculation of all components of $\varepsilon_c$. 

\begin{figure}[b]
    \centering
    \includegraphics[width = 1\columnwidth]{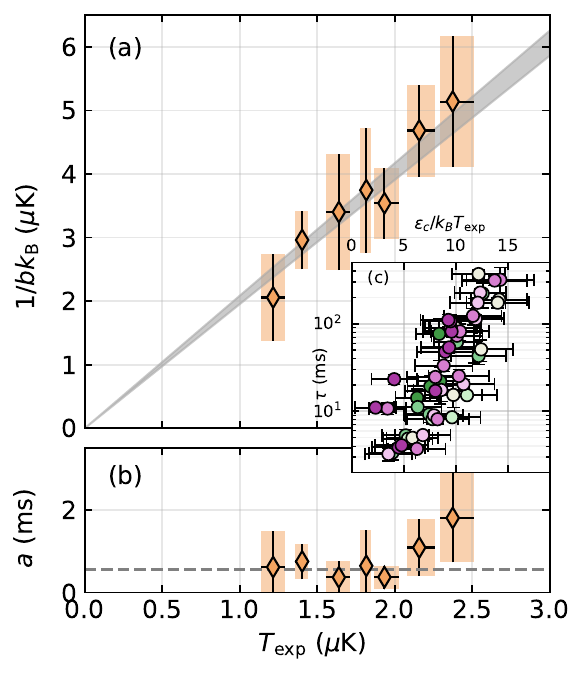}
        \caption{(a) The experimentally measured (diamonds) values of $1/bk_B$ as a function of the atomic temperature $T_\mathrm{exp}$ show a linear dependence. 
        Orange boxes are horizontally bounded by the minimum and maximum temperature in each cluster, while vertical boundaries result from the fit uncertainties on $b$. Points are centered around the mean value of $T_\mathrm{exp}$ for each cluster. The shaded grey area corresponds to the one-$\sigma$ confidence interval of a linear fit. (b) Within the experimental uncertainties, the prefactor $a$ does not show any clear dependence on $T_\mathrm{exp}$ and has a mean value of 0.8(4)\,ms. (c) The $\tau$ data from Fig.~\ref{fig:fig3} collapse onto an single curve when plotted as a function of the dimensionless parameter $\varepsilon_c/k_BT_\mathrm{exp}$.}
    \label{fig:fig4}
\end{figure}

The experimental values of $\tau$ from each temperature cluster are fitted as a function of $\varepsilon_c$ with the following equation, formally equivalent to Eq.~\ref{eq:eq1}
\begin{equation}
   \ln(\tau)= b\,{\varepsilon_c}+\ln (a)-\frac12\ln(b\,{\varepsilon_c}),
    \label{eq:tau-fit}
\end{equation}
with $a$ and $b$ the only two fitting parameters. 
Ideally, this procedure allows us to find the temperature dependence of $a$ and $b$.

Examples of experimentally obtained $\tau$ and relative fits are reported in \aref{fig:fig3}(b) for two different values of the sample temperature. More information on the analysis procedure can be found in~\cite{SupMat}. In lin-log scale the data and the fit curves show a linear behavior for large $\varepsilon_c/k_B$, with decreasing slope for higher $T_\mathrm{exp}$. This agrees with  Eq.\,\ref{eq:eq1} when the exponential term dominates over the denominator, i.e., for large $\beta E_c$.
The quantity $1/bk_B$ resulting from the fits is presented as a function of the measured temperature $T_\mathrm{exp}$ in \aref{fig:fig4}(a). The data suggest a direct proportionality between the two, confirming the theoretical prediction that the thermally induced FVD rate is dominated by an exponential growth with the system temperature, as expressed by Eq.\,\ref{eq:eq1}. A linear fit provides $1/bk_B=2.02(7)\times T_\mathrm{exp}$.

Figure~\ref{fig:fig4}(b) highlights that in the measured temperature range, $a$ is compatible within experimental uncertainties to a constant value of 0.8(4)\,ms.  Thanks to this, the decay times $\tau$ data collapse together when plotted as a function of $\varepsilon_c/k_B T_\mathrm{exp}$, as visible in \aref{fig:fig4}(c).
The robustness of our analysis is confirmed by the persisting linear dependence also when the number of clusters in which the data are grouped is increased or decreased. 

In conclusion, we measured the temperature dependence of the FVD rate on a coherently-coupled quantum mixture with ferromagnetic properties, finding good agreement with the extension of the instanton theory to the finite temperature case. In particular, we proved that our tunable experimental platform allows to explore field theories at finite temperature as for instance the Higgs field stability.
These results provide new material to understand FVD and instanton theory applied to atomic systems and trigger the experimental and theoretical investigation of the decay mechanism also in non-scalar fields and of the effects of dissipation on $\tau$ \cite{Pirvu2024, Maki2023}. 
The possibility to observe the FVD in the purely quantum regime remains challenging and connected to a direct thermometry of the spin channel, for example by applying fluctuation-dissipation theorem on spin and density correlations \cite{Abad2013}.\\

\textit{Acknowledgments}--We thank Anna Berti, Iacopo Carusotto, Alessio Recati, and Ian Moss for fruitful discussions.
We acknowledge funding from Provincia Autonoma di Trento, from INFN through the RELAQS project,
from the European Union’s Horizon 2020 research and innovation Programme through the STAQS project of QuantERA II (Grant Agreement No. 101017733), and from the European Union - Next Generation EU through PNRR MUR project PE0000023-NQSTI. We also acknowledge the project DYNAMITE QUANTERA2-00056 funded by the Ministry of University and Research through the ERANET COFUND QuantERA II – 2021 call and co-funded by the European Union (H2020, Grant Agreement No. 101017733). This work was supported by Q@TN, the joint lab between the University of Trento, FBK - Fondazione Bruno Kessler, INFN - National Institute for Nuclear Physics and CNR - National Research Council.\\

\textit{Data availability}--The data that support the findings of
this article are openly available \cite{Data}.


\begin{thebibliography}{40}%
\makeatletter
\providecommand \@ifxundefined [1]{%
 \@ifx{#1\undefined}
}%
\providecommand \@ifnum [1]{%
 \ifnum #1\expandafter \@firstoftwo
 \else \expandafter \@secondoftwo
 \fi
}%
\providecommand \@ifx [1]{%
 \ifx #1\expandafter \@firstoftwo
 \else \expandafter \@secondoftwo
 \fi
}%
\providecommand \natexlab [1]{#1}%
\providecommand \enquote  [1]{``#1''}%
\providecommand \bibnamefont  [1]{#1}%
\providecommand \bibfnamefont [1]{#1}%
\providecommand \citenamefont [1]{#1}%
\providecommand \href@noop [0]{\@secondoftwo}%
\providecommand \href [0]{\begingroup \@sanitize@url \@href}%
\providecommand \@href[1]{\@@startlink{#1}\@@href}%
\providecommand \@@href[1]{\endgroup#1\@@endlink}%
\providecommand \@sanitize@url [0]{\catcode `\\12\catcode `\$12\catcode `\&12\catcode `\#12\catcode `\^12\catcode `\_12\catcode `\%12\relax}%
\providecommand \@@startlink[1]{}%
\providecommand \@@endlink[0]{}%
\providecommand \url  [0]{\begingroup\@sanitize@url \@url }%
\providecommand \@url [1]{\endgroup\@href {#1}{\urlprefix }}%
\providecommand \urlprefix  [0]{URL }%
\providecommand \Eprint [0]{\href }%
\providecommand \doibase [0]{https://doi.org/}%
\providecommand \selectlanguage [0]{\@gobble}%
\providecommand \bibinfo  [0]{\@secondoftwo}%
\providecommand \bibfield  [0]{\@secondoftwo}%
\providecommand \translation [1]{[#1]}%
\providecommand \BibitemOpen [0]{}%
\providecommand \bibitemStop [0]{}%
\providecommand \bibitemNoStop [0]{.\EOS\space}%
\providecommand \EOS [0]{\spacefactor3000\relax}%
\providecommand \BibitemShut  [1]{\csname bibitem#1\endcsname}%
\let\auto@bib@innerbib\@empty
\bibitem [{\citenamefont {Franks}(1982)}]{Franks1982}%
  \BibitemOpen
  \bibfield  {author} {\bibinfo {author} {\bibfnamefont {F.}~\bibnamefont {Franks}},\ }\href {https://doi.org/doi.org/10.1007/978-1-4757-6952-4} {\emph {\bibinfo {title} {{Water and Aqueous Solutions at Subzero Temperatures}}}}\ (\bibinfo  {publisher} {Springe},\ \bibinfo {year} {1982})\BibitemShut {NoStop}%
\bibitem [{\citenamefont {Devoto}\ \emph {et~al.}(2022)\citenamefont {Devoto}, \citenamefont {Devoto}, \citenamefont {Luzio},\ and\ \citenamefont {Ridolfi}}]{Devoto2022}%
  \BibitemOpen
  \bibfield  {author} {\bibinfo {author} {\bibfnamefont {F.}~\bibnamefont {Devoto}}, \bibinfo {author} {\bibfnamefont {S.}~\bibnamefont {Devoto}}, \bibinfo {author} {\bibfnamefont {L.~D.}\ \bibnamefont {Luzio}},\ and\ \bibinfo {author} {\bibfnamefont {G.}~\bibnamefont {Ridolfi}},\ }\bibfield  {title} {\bibinfo {title} {False vacuum decay: an introductory review},\ }\href {https://doi.org/10.1088/1361-6471/ac7f24} {\bibfield  {journal} {\bibinfo  {journal} {Journal of Physics G: Nuclear and Particle Physics}\ }\textbf {\bibinfo {volume} {49}},\ \bibinfo {pages} {103001} (\bibinfo {year} {2022})}\BibitemShut {NoStop}%
\bibitem [{\citenamefont {DeVries}\ and\ \citenamefont {Wohlschlag}(1969)}]{devries1969}%
  \BibitemOpen
  \bibfield  {author} {\bibinfo {author} {\bibfnamefont {A.~L.}\ \bibnamefont {DeVries}}\ and\ \bibinfo {author} {\bibfnamefont {D.~E.}\ \bibnamefont {Wohlschlag}},\ }\bibfield  {title} {\bibinfo {title} {Freezing resistance in some antarctic fishes},\ }\href {https://doi.org/10.1126/science.163.3871.1073} {\bibfield  {journal} {\bibinfo  {journal} {Science}\ }\textbf {\bibinfo {volume} {163}},\ \bibinfo {pages} {1073} (\bibinfo {year} {1969})}\BibitemShut {NoStop}%
\bibitem [{\citenamefont {Herlach}(2001)}]{herlach2001}%
  \BibitemOpen
  \bibfield  {author} {\bibinfo {author} {\bibfnamefont {D.~M.}\ \bibnamefont {Herlach}},\ }\bibfield  {title} {\bibinfo {title} {Metastable materials solidified from undercooled melts},\ }\href {https://doi.org/10.1088/0953-8984/13/34/317} {\bibfield  {journal} {\bibinfo  {journal} {Journal of Physics: Condensed Matter}\ }\textbf {\bibinfo {volume} {13}},\ \bibinfo {pages} {7737} (\bibinfo {year} {2001})}\BibitemShut {NoStop}%
\bibitem [{\citenamefont {Gallo}\ \emph {et~al.}(2021)\citenamefont {Gallo}, \citenamefont {Magaletti},\ and\ \citenamefont {Casciola}}]{Gallo2021}%
  \BibitemOpen
  \bibfield  {author} {\bibinfo {author} {\bibfnamefont {M.}~\bibnamefont {Gallo}}, \bibinfo {author} {\bibfnamefont {F.}~\bibnamefont {Magaletti}},\ and\ \bibinfo {author} {\bibfnamefont {C.~M.}\ \bibnamefont {Casciola}},\ }\bibfield  {title} {\bibinfo {title} {Heterogeneous bubble nucleation dynamics},\ }\href {https://doi.org/10.1017/jfm.2020.761} {\bibfield  {journal} {\bibinfo  {journal} {Journal of Fluid Mechanics}\ }\textbf {\bibinfo {volume} {906}},\ \bibinfo {pages} {A20} (\bibinfo {year} {2021})}\BibitemShut {NoStop}%
\bibitem [{\citenamefont {Kulmala}\ \emph {et~al.}(2013)\citenamefont {Kulmala}, \citenamefont {Kontkanen}, \citenamefont {Junninen}, \citenamefont {Lehtipalo}, \citenamefont {Manninen}, \citenamefont {Nieminen}, \citenamefont {Petäjä}, \citenamefont {Sipilä}, \citenamefont {Schobesberger}, \citenamefont {Rantala}, \citenamefont {Franchin}, \citenamefont {Jokinen}, \citenamefont {Järvinen}, \citenamefont {Äijälä}, \citenamefont {Kangasluoma}, \citenamefont {Hakala}, \citenamefont {Aalto}, \citenamefont {Paasonen}, \citenamefont {Mikkilä}, \citenamefont {Vanhanen}, \citenamefont {Aalto}, \citenamefont {Hakola}, \citenamefont {Makkonen}, \citenamefont {Ruuskanen}, \citenamefont {Mauldin}, \citenamefont {Duplissy}, \citenamefont {Vehkamäki}, \citenamefont {Bäck}, \citenamefont {Kortelainen}, \citenamefont {Riipinen}, \citenamefont {Kurtén}, \citenamefont {Johnston}, \citenamefont {Smith}, \citenamefont {Ehn}, \citenamefont {Mentel}, \citenamefont {Lehtinen}, \citenamefont {Laaksonen}, \citenamefont
  {Kerminen},\ and\ \citenamefont {Worsnop}}]{Kulmala2013}%
  \BibitemOpen
  \bibfield  {author} {\bibinfo {author} {\bibfnamefont {M.}~\bibnamefont {Kulmala}}, \bibinfo {author} {\bibfnamefont {J.}~\bibnamefont {Kontkanen}}, \bibinfo {author} {\bibfnamefont {H.}~\bibnamefont {Junninen}}, \bibinfo {author} {\bibfnamefont {K.}~\bibnamefont {Lehtipalo}}, \bibinfo {author} {\bibfnamefont {H.~E.}\ \bibnamefont {Manninen}}, \bibinfo {author} {\bibfnamefont {T.}~\bibnamefont {Nieminen}}, \bibinfo {author} {\bibfnamefont {T.}~\bibnamefont {Petäjä}}, \bibinfo {author} {\bibfnamefont {M.}~\bibnamefont {Sipilä}}, \bibinfo {author} {\bibfnamefont {S.}~\bibnamefont {Schobesberger}}, \bibinfo {author} {\bibfnamefont {P.}~\bibnamefont {Rantala}}, \bibinfo {author} {\bibfnamefont {A.}~\bibnamefont {Franchin}}, \bibinfo {author} {\bibfnamefont {T.}~\bibnamefont {Jokinen}}, \bibinfo {author} {\bibfnamefont {E.}~\bibnamefont {Järvinen}}, \bibinfo {author} {\bibfnamefont {M.}~\bibnamefont {Äijälä}}, \bibinfo {author} {\bibfnamefont {J.}~\bibnamefont {Kangasluoma}}, \bibinfo {author}
  {\bibfnamefont {J.}~\bibnamefont {Hakala}}, \bibinfo {author} {\bibfnamefont {P.~P.}\ \bibnamefont {Aalto}}, \bibinfo {author} {\bibfnamefont {P.}~\bibnamefont {Paasonen}}, \bibinfo {author} {\bibfnamefont {J.}~\bibnamefont {Mikkilä}}, \bibinfo {author} {\bibfnamefont {J.}~\bibnamefont {Vanhanen}}, \bibinfo {author} {\bibfnamefont {J.}~\bibnamefont {Aalto}}, \bibinfo {author} {\bibfnamefont {H.}~\bibnamefont {Hakola}}, \bibinfo {author} {\bibfnamefont {U.}~\bibnamefont {Makkonen}}, \bibinfo {author} {\bibfnamefont {T.}~\bibnamefont {Ruuskanen}}, \bibinfo {author} {\bibfnamefont {R.~L.}\ \bibnamefont {Mauldin}}, \bibinfo {author} {\bibfnamefont {J.}~\bibnamefont {Duplissy}}, \bibinfo {author} {\bibfnamefont {H.}~\bibnamefont {Vehkamäki}}, \bibinfo {author} {\bibfnamefont {J.}~\bibnamefont {Bäck}}, \bibinfo {author} {\bibfnamefont {A.}~\bibnamefont {Kortelainen}}, \bibinfo {author} {\bibfnamefont {I.}~\bibnamefont {Riipinen}}, \bibinfo {author} {\bibfnamefont {T.}~\bibnamefont {Kurtén}}, \bibinfo {author}
  {\bibfnamefont {M.~V.}\ \bibnamefont {Johnston}}, \bibinfo {author} {\bibfnamefont {J.~N.}\ \bibnamefont {Smith}}, \bibinfo {author} {\bibfnamefont {M.}~\bibnamefont {Ehn}}, \bibinfo {author} {\bibfnamefont {T.~F.}\ \bibnamefont {Mentel}}, \bibinfo {author} {\bibfnamefont {K.~E.~J.}\ \bibnamefont {Lehtinen}}, \bibinfo {author} {\bibfnamefont {A.}~\bibnamefont {Laaksonen}}, \bibinfo {author} {\bibfnamefont {V.-M.}\ \bibnamefont {Kerminen}},\ and\ \bibinfo {author} {\bibfnamefont {D.~R.}\ \bibnamefont {Worsnop}},\ }\bibfield  {title} {\bibinfo {title} {Direct observations of atmospheric aerosol nucleation},\ }\href {https://doi.org/10.1126/science.1227385} {\bibfield  {journal} {\bibinfo  {journal} {Science}\ }\textbf {\bibinfo {volume} {339}},\ \bibinfo {pages} {943} (\bibinfo {year} {2013})}\BibitemShut {NoStop}%
\bibitem [{\citenamefont {Sachdev}(2011)}]{Sachdev2011}%
  \BibitemOpen
  \bibfield  {author} {\bibinfo {author} {\bibfnamefont {S.}~\bibnamefont {Sachdev}},\ }\href {https://doi.org/doi.org/10.1017/CBO9780511973765} {\emph {\bibinfo {title} {Quantum phase transitions}}}\ (\bibinfo  {publisher} {Cambridge University Press},\ \bibinfo {address} {New York},\ \bibinfo {year} {2011})\ Chap.~\bibinfo {chapter} {16}\BibitemShut {NoStop}%
\bibitem [{\citenamefont {Kobzarev}\ \emph {et~al.}(1974)\citenamefont {Kobzarev}, \citenamefont {Okun},\ and\ \citenamefont {Voloshin}}]{Kobzarev1974}%
  \BibitemOpen
  \bibfield  {author} {\bibinfo {author} {\bibfnamefont {I.~Y.}\ \bibnamefont {Kobzarev}}, \bibinfo {author} {\bibfnamefont {L.~B.}\ \bibnamefont {Okun}},\ and\ \bibinfo {author} {\bibfnamefont {M.~B.}\ \bibnamefont {Voloshin}},\ }\bibfield  {title} {\bibinfo {title} {{Bubbles in Metastable Vacuum}},\ }\href@noop {} {\bibfield  {journal} {\bibinfo  {journal} {Yad. Fiz.}\ }\textbf {\bibinfo {volume} {20}},\ \bibinfo {pages} {1229} (\bibinfo {year} {1974})}\BibitemShut {NoStop}%
\bibitem [{\citenamefont {Coleman}(1977{\natexlab{a}})}]{Coleman1977}%
  \BibitemOpen
  \bibfield  {author} {\bibinfo {author} {\bibfnamefont {S.}~\bibnamefont {Coleman}},\ }\bibfield  {title} {\bibinfo {title} {Fate of the false vacuum: Semiclassical theory},\ }\href {https://doi.org/10.1103/PhysRevD.15.2929} {\bibfield  {journal} {\bibinfo  {journal} {Phys. Rev. D}\ }\textbf {\bibinfo {volume} {15}},\ \bibinfo {pages} {2929} (\bibinfo {year} {1977}{\natexlab{a}})}\BibitemShut {NoStop}%
\bibitem [{\citenamefont {Coleman}(1977{\natexlab{b}})}]{Coleman1977erratum}%
  \BibitemOpen
  \bibfield  {author} {\bibinfo {author} {\bibfnamefont {S.~R.}\ \bibnamefont {Coleman}},\ }\bibfield  {title} {\bibinfo {title} {{The Fate of the False Vacuum. 1. Semiclassical Theory}},\ }\href {https://doi.org/10.1103/PhysRevD.15.2929, 10.1103/PhysRevD.16.1248} {\bibfield  {journal} {\bibinfo  {journal} {Phys. Rev. D}\ }\textbf {\bibinfo {volume} {15}},\ \bibinfo {pages} {2929} (\bibinfo {year} {1977}{\natexlab{b}})},\ \bibinfo {note} {[Erratum: Phys. Rev. D 16, 1248 (1977)]}\BibitemShut {NoStop}%
\bibitem [{\citenamefont {Callan}\ and\ \citenamefont {Coleman}(1977)}]{Callan1977}%
  \BibitemOpen
  \bibfield  {author} {\bibinfo {author} {\bibfnamefont {C.~G.}\ \bibnamefont {Callan}}\ and\ \bibinfo {author} {\bibfnamefont {S.}~\bibnamefont {Coleman}},\ }\bibfield  {title} {\bibinfo {title} {Fate of the false vacuum. ii. first quantum corrections},\ }\href {https://doi.org/10.1103/PhysRevD.16.1762} {\bibfield  {journal} {\bibinfo  {journal} {Phys. Rev. D}\ }\textbf {\bibinfo {volume} {16}},\ \bibinfo {pages} {1762} (\bibinfo {year} {1977})}\BibitemShut {NoStop}%
\bibitem [{\citenamefont {Langer}(1969)}]{LANGER1969}%
  \BibitemOpen
  \bibfield  {author} {\bibinfo {author} {\bibfnamefont {J.}~\bibnamefont {Langer}},\ }\bibfield  {title} {\bibinfo {title} {Statistical theory of the decay of metastable states},\ }\href {https://doi.org/10.1016/0003-4916(69)90153-5} {\bibfield  {journal} {\bibinfo  {journal} {Annals of Physics}\ }\textbf {\bibinfo {volume} {54}},\ \bibinfo {pages} {258} (\bibinfo {year} {1969})}\BibitemShut {NoStop}%
\bibitem [{\citenamefont {Quiros}(1999)}]{Quiros1999}%
  \BibitemOpen
  \bibfield  {author} {\bibinfo {author} {\bibfnamefont {M.}~\bibnamefont {Quiros}},\ }\href@noop {} {\bibinfo {title} {Finite temperature field theory and phase transitions}} (\bibinfo {year} {1999}),\ \Eprint {https://arxiv.org/abs/hep-ph/9901312} {arXiv:hep-ph/9901312 [hep-ph]} \BibitemShut {NoStop}%
\bibitem [{\citenamefont {Fialko}\ \emph {et~al.}(2015)\citenamefont {Fialko}, \citenamefont {Opanchuk}, \citenamefont {Sidorov}, \citenamefont {Drummond},\ and\ \citenamefont {Brand}}]{Fialko2015}%
  \BibitemOpen
  \bibfield  {author} {\bibinfo {author} {\bibfnamefont {O.}~\bibnamefont {Fialko}}, \bibinfo {author} {\bibfnamefont {B.}~\bibnamefont {Opanchuk}}, \bibinfo {author} {\bibfnamefont {A.~I.}\ \bibnamefont {Sidorov}}, \bibinfo {author} {\bibfnamefont {P.~D.}\ \bibnamefont {Drummond}},\ and\ \bibinfo {author} {\bibfnamefont {J.}~\bibnamefont {Brand}},\ }\bibfield  {title} {\bibinfo {title} {Fate of the false vacuum: Towards realization with ultra-cold atoms},\ }\href {https://doi.org/10.1209/0295-5075/110/56001} {\bibfield  {journal} {\bibinfo  {journal} {Europhysics Letters}\ }\textbf {\bibinfo {volume} {110}},\ \bibinfo {pages} {56001} (\bibinfo {year} {2015})}\BibitemShut {NoStop}%
\bibitem [{\citenamefont {Braden}\ \emph {et~al.}(2019)\citenamefont {Braden}, \citenamefont {Johnson}, \citenamefont {Peiris},\ and\ \citenamefont {Weinfurtner}}]{Braden2018}%
  \BibitemOpen
  \bibfield  {author} {\bibinfo {author} {\bibfnamefont {J.}~\bibnamefont {Braden}}, \bibinfo {author} {\bibfnamefont {M.~C.}\ \bibnamefont {Johnson}}, \bibinfo {author} {\bibfnamefont {H.~V.}\ \bibnamefont {Peiris}},\ and\ \bibinfo {author} {\bibfnamefont {S.}~\bibnamefont {Weinfurtner}},\ }\bibfield  {title} {\bibinfo {title} {Towards the cold atom analog false vacuum},\ }\href {https://doi.org/10.1007/JHEP07(2018)014} {\bibfield  {journal} {\bibinfo  {journal} {Journal of High Energy Physics}\ }\textbf {\bibinfo {volume} {2018}},\ \bibinfo {pages} {2018} (\bibinfo {year} {2019})}\BibitemShut {NoStop}%
\bibitem [{\citenamefont {Billam}\ \emph {et~al.}(2019)\citenamefont {Billam}, \citenamefont {Gregory}, \citenamefont {Michel},\ and\ \citenamefont {Moss}}]{Billam2019}%
  \BibitemOpen
  \bibfield  {author} {\bibinfo {author} {\bibfnamefont {T.~P.}\ \bibnamefont {Billam}}, \bibinfo {author} {\bibfnamefont {R.}~\bibnamefont {Gregory}}, \bibinfo {author} {\bibfnamefont {F.}~\bibnamefont {Michel}},\ and\ \bibinfo {author} {\bibfnamefont {I.~G.}\ \bibnamefont {Moss}},\ }\bibfield  {title} {\bibinfo {title} {Simulating seeded vacuum decay in a cold atom system},\ }\href {https://doi.org/10.1103/PhysRevD.100.065016} {\bibfield  {journal} {\bibinfo  {journal} {Phys. Rev. D}\ }\textbf {\bibinfo {volume} {100}},\ \bibinfo {pages} {065016} (\bibinfo {year} {2019})}\BibitemShut {NoStop}%
\bibitem [{\citenamefont {Billam}\ \emph {et~al.}(2020)\citenamefont {Billam}, \citenamefont {Brown},\ and\ \citenamefont {Moss}}]{Billam2020}%
  \BibitemOpen
  \bibfield  {author} {\bibinfo {author} {\bibfnamefont {T.~P.}\ \bibnamefont {Billam}}, \bibinfo {author} {\bibfnamefont {K.}~\bibnamefont {Brown}},\ and\ \bibinfo {author} {\bibfnamefont {I.~G.}\ \bibnamefont {Moss}},\ }\bibfield  {title} {\bibinfo {title} {Simulating cosmological supercooling with a cold-atom system},\ }\href {https://doi.org/10.1103/PhysRevA.102.043324} {\bibfield  {journal} {\bibinfo  {journal} {Phys. Rev. A}\ }\textbf {\bibinfo {volume} {102}},\ \bibinfo {pages} {043324} (\bibinfo {year} {2020})}\BibitemShut {NoStop}%
\bibitem [{\citenamefont {Billam}\ \emph {et~al.}(2021)\citenamefont {Billam}, \citenamefont {Brown}, \citenamefont {Groszek},\ and\ \citenamefont {Moss}}]{Billam2021}%
  \BibitemOpen
  \bibfield  {author} {\bibinfo {author} {\bibfnamefont {T.~P.}\ \bibnamefont {Billam}}, \bibinfo {author} {\bibfnamefont {K.}~\bibnamefont {Brown}}, \bibinfo {author} {\bibfnamefont {A.~J.}\ \bibnamefont {Groszek}},\ and\ \bibinfo {author} {\bibfnamefont {I.~G.}\ \bibnamefont {Moss}},\ }\bibfield  {title} {\bibinfo {title} {Simulating cosmological supercooling with a cold atom system. ii. thermal damping and parametric instability},\ }\href {https://doi.org/10.1103/PhysRevA.104.053309} {\bibfield  {journal} {\bibinfo  {journal} {Phys. Rev. A}\ }\textbf {\bibinfo {volume} {104}},\ \bibinfo {pages} {053309} (\bibinfo {year} {2021})}\BibitemShut {NoStop}%
\bibitem [{\citenamefont {Jenkins}\ \emph {et~al.}(2024)\citenamefont {Jenkins}, \citenamefont {Braden}, \citenamefont {Peiris}, \citenamefont {Pontzen}, \citenamefont {Johnson},\ and\ \citenamefont {Weinfurtner}}]{Jenkins2024}%
  \BibitemOpen
  \bibfield  {author} {\bibinfo {author} {\bibfnamefont {A.~C.}\ \bibnamefont {Jenkins}}, \bibinfo {author} {\bibfnamefont {J.}~\bibnamefont {Braden}}, \bibinfo {author} {\bibfnamefont {H.~V.}\ \bibnamefont {Peiris}}, \bibinfo {author} {\bibfnamefont {A.}~\bibnamefont {Pontzen}}, \bibinfo {author} {\bibfnamefont {M.~C.}\ \bibnamefont {Johnson}},\ and\ \bibinfo {author} {\bibfnamefont {S.}~\bibnamefont {Weinfurtner}},\ }\bibfield  {title} {\bibinfo {title} {Analog vacuum decay from vacuum initial conditions},\ }\href {https://doi.org/10.1103/PhysRevD.109.023506} {\bibfield  {journal} {\bibinfo  {journal} {Phys. Rev. D}\ }\textbf {\bibinfo {volume} {109}},\ \bibinfo {pages} {023506} (\bibinfo {year} {2024})}\BibitemShut {NoStop}%
\bibitem [{\citenamefont {Mazumdar}\ and\ \citenamefont {White}(2019)}]{Mazumdar2019}%
  \BibitemOpen
  \bibfield  {author} {\bibinfo {author} {\bibfnamefont {A.}~\bibnamefont {Mazumdar}}\ and\ \bibinfo {author} {\bibfnamefont {G.}~\bibnamefont {White}},\ }\bibfield  {title} {\bibinfo {title} {Review of cosmic phase transitions: their significance and experimental signatures},\ }\href {https://doi.org/10.1088/1361-6633/ab1f55} {\bibfield  {journal} {\bibinfo  {journal} {Reports on Progress in Physics}\ }\textbf {\bibinfo {volume} {82}},\ \bibinfo {pages} {076901} (\bibinfo {year} {2019})}\BibitemShut {NoStop}%
\bibitem [{\citenamefont {Hindmarsh}\ \emph {et~al.}(2021)\citenamefont {Hindmarsh}, \citenamefont {Lüben}, \citenamefont {Lumma},\ and\ \citenamefont {Pauly}}]{Hindmarsh2021}%
  \BibitemOpen
  \bibfield  {author} {\bibinfo {author} {\bibfnamefont {M.}~\bibnamefont {Hindmarsh}}, \bibinfo {author} {\bibfnamefont {M.}~\bibnamefont {Lüben}}, \bibinfo {author} {\bibfnamefont {J.}~\bibnamefont {Lumma}},\ and\ \bibinfo {author} {\bibfnamefont {M.}~\bibnamefont {Pauly}},\ }\bibfield  {title} {\bibinfo {title} {{Phase transitions in the early universe}},\ }\href {https://doi.org/10.21468/SciPostPhysLectNotes.24} {\bibfield  {journal} {\bibinfo  {journal} {SciPost Phys. Lect. Notes}\ ,\ \bibinfo {pages} {24}} (\bibinfo {year} {2021})}\BibitemShut {NoStop}%
\bibitem [{\citenamefont {{Lifshitz}}\ and\ \citenamefont {{Kagan}}(1972)}]{Lifshitz72}%
  \BibitemOpen
  \bibfield  {author} {\bibinfo {author} {\bibfnamefont {I.~M.}\ \bibnamefont {{Lifshitz}}}\ and\ \bibinfo {author} {\bibfnamefont {Y.}~\bibnamefont {{Kagan}}},\ }\bibfield  {title} {\bibinfo {title} {{Quantum Kinetics of Phase Transitions at Temperatures Close to Absolute Zero}},\ }\href@noop {} {\bibfield  {journal} {\bibinfo  {journal} {Soviet Journal of Experimental and Theoretical Physics}\ }\textbf {\bibinfo {volume} {35}},\ \bibinfo {pages} {206} (\bibinfo {year} {1972})}\BibitemShut {NoStop}%
\bibitem [{\citenamefont {Gorokhov}\ and\ \citenamefont {Blatter}(1997)}]{Gorokhov97}%
  \BibitemOpen
  \bibfield  {author} {\bibinfo {author} {\bibfnamefont {D.~A.}\ \bibnamefont {Gorokhov}}\ and\ \bibinfo {author} {\bibfnamefont {G.}~\bibnamefont {Blatter}},\ }\bibfield  {title} {\bibinfo {title} {Decay of metastable states: Sharp transition from quantum to classical behavior},\ }\href {https://doi.org/10.1103/PhysRevB.56.3130} {\bibfield  {journal} {\bibinfo  {journal} {Phys. Rev. B}\ }\textbf {\bibinfo {volume} {56}},\ \bibinfo {pages} {3130} (\bibinfo {year} {1997})}\BibitemShut {NoStop}%
\bibitem [{\citenamefont {Linde}(1983)}]{Linde1983}%
  \BibitemOpen
  \bibfield  {author} {\bibinfo {author} {\bibfnamefont {A.}~\bibnamefont {Linde}},\ }\bibfield  {title} {\bibinfo {title} {Decay of the false vacuum at finite temperature},\ }\href {https://doi.org/https://doi.org/10.1016/0550-3213(83)90293-6} {\bibfield  {journal} {\bibinfo  {journal} {Nuclear Physics B}\ }\textbf {\bibinfo {volume} {216}},\ \bibinfo {pages} {421} (\bibinfo {year} {1983})}\BibitemShut {NoStop}%
\bibitem [{\citenamefont {Vodeb}\ \emph {et~al.}(2025)\citenamefont {Vodeb}, \citenamefont {Desaules}, \citenamefont {Hallam}, \citenamefont {Rava}, \citenamefont {Humar}, \citenamefont {Willsch}, \citenamefont {Jin}, \citenamefont {Willsch}, \citenamefont {Michielsen},\ and\ \citenamefont {Papi{\'c}}}]{Vodeb2025}%
  \BibitemOpen
  \bibfield  {author} {\bibinfo {author} {\bibfnamefont {J.}~\bibnamefont {Vodeb}}, \bibinfo {author} {\bibfnamefont {J.-Y.}\ \bibnamefont {Desaules}}, \bibinfo {author} {\bibfnamefont {A.}~\bibnamefont {Hallam}}, \bibinfo {author} {\bibfnamefont {A.}~\bibnamefont {Rava}}, \bibinfo {author} {\bibfnamefont {G.}~\bibnamefont {Humar}}, \bibinfo {author} {\bibfnamefont {D.}~\bibnamefont {Willsch}}, \bibinfo {author} {\bibfnamefont {F.}~\bibnamefont {Jin}}, \bibinfo {author} {\bibfnamefont {M.}~\bibnamefont {Willsch}}, \bibinfo {author} {\bibfnamefont {K.}~\bibnamefont {Michielsen}},\ and\ \bibinfo {author} {\bibfnamefont {Z.}~\bibnamefont {Papi{\'c}}},\ }\bibfield  {title} {\bibinfo {title} {Stirring the false vacuum via interacting quantized bubbles on a 5,564-qubit quantum annealer},\ }\href {https://doi.org/10.1038/s41567-024-02765-w} {\bibfield  {journal} {\bibinfo  {journal} {Nature Physics}\ }\textbf {\bibinfo {volume} {21}},\ \bibinfo {pages} {386} (\bibinfo {year} {2025})}\BibitemShut {NoStop}%
\bibitem [{\citenamefont {Zhu}\ \emph {et~al.}(2024)\citenamefont {Zhu}, \citenamefont {Liu}, \citenamefont {Lagnese}, \citenamefont {Surace}, \citenamefont {Zhang}, \citenamefont {He}, \citenamefont {Halimeh}, \citenamefont {Dalmonte}, \citenamefont {Morampudi}, \citenamefont {Wilczek}, \citenamefont {Yuan},\ and\ \citenamefont {Pan}}]{Zhu2024}%
  \BibitemOpen
  \bibfield  {author} {\bibinfo {author} {\bibfnamefont {Z.-H.}\ \bibnamefont {Zhu}}, \bibinfo {author} {\bibfnamefont {Y.}~\bibnamefont {Liu}}, \bibinfo {author} {\bibfnamefont {G.}~\bibnamefont {Lagnese}}, \bibinfo {author} {\bibfnamefont {F.~M.}\ \bibnamefont {Surace}}, \bibinfo {author} {\bibfnamefont {W.-Y.}\ \bibnamefont {Zhang}}, \bibinfo {author} {\bibfnamefont {M.-G.}\ \bibnamefont {He}}, \bibinfo {author} {\bibfnamefont {J.~C.}\ \bibnamefont {Halimeh}}, \bibinfo {author} {\bibfnamefont {M.}~\bibnamefont {Dalmonte}}, \bibinfo {author} {\bibfnamefont {S.~C.}\ \bibnamefont {Morampudi}}, \bibinfo {author} {\bibfnamefont {F.}~\bibnamefont {Wilczek}}, \bibinfo {author} {\bibfnamefont {Z.-S.}\ \bibnamefont {Yuan}},\ and\ \bibinfo {author} {\bibfnamefont {J.-W.}\ \bibnamefont {Pan}},\ }\href@noop {} {\bibinfo {title} {Probing false vacuum decay on a cold-atom gauge-theory quantum simulator}} (\bibinfo {year} {2024}),\ \Eprint {https://arxiv.org/abs/2411.12565} {arXiv:2411.12565 [cond-mat.quant-gas]}
  \BibitemShut {NoStop}%
\bibitem [{\citenamefont {Luo}\ \emph {et~al.}(2025)\citenamefont {Luo}, \citenamefont {Surace}, \citenamefont {De}, \citenamefont {Lerose}, \citenamefont {Bennewitz}, \citenamefont {Ware}, \citenamefont {Schuckert}, \citenamefont {Davoudi}, \citenamefont {Gorshkov}, \citenamefont {Katz},\ and\ \citenamefont {Monroe}}]{Luo2025}%
  \BibitemOpen
  \bibfield  {author} {\bibinfo {author} {\bibfnamefont {D.}~\bibnamefont {Luo}}, \bibinfo {author} {\bibfnamefont {F.~M.}\ \bibnamefont {Surace}}, \bibinfo {author} {\bibfnamefont {A.}~\bibnamefont {De}}, \bibinfo {author} {\bibfnamefont {A.}~\bibnamefont {Lerose}}, \bibinfo {author} {\bibfnamefont {E.~R.}\ \bibnamefont {Bennewitz}}, \bibinfo {author} {\bibfnamefont {B.}~\bibnamefont {Ware}}, \bibinfo {author} {\bibfnamefont {A.}~\bibnamefont {Schuckert}}, \bibinfo {author} {\bibfnamefont {Z.}~\bibnamefont {Davoudi}}, \bibinfo {author} {\bibfnamefont {A.~V.}\ \bibnamefont {Gorshkov}}, \bibinfo {author} {\bibfnamefont {O.}~\bibnamefont {Katz}},\ and\ \bibinfo {author} {\bibfnamefont {C.}~\bibnamefont {Monroe}},\ }\href@noop {} {\bibinfo {title} {Quantum simulation of bubble nucleation across a quantum phase transition}} (\bibinfo {year} {2025}),\ \Eprint {https://arxiv.org/abs/2505.09607} {arXiv:2505.09607 [quant-ph]} \BibitemShut {NoStop}%
\bibitem [{\citenamefont {Zenesini}\ \emph {et~al.}(2024)\citenamefont {Zenesini}, \citenamefont {Berti}, \citenamefont {Cominotti}, \citenamefont {Rogora}, \citenamefont {Moss}, \citenamefont {Billam}, \citenamefont {Carusotto}, \citenamefont {Lamporesi}, \citenamefont {Recati},\ and\ \citenamefont {Ferrari}}]{Zenesini2024}%
  \BibitemOpen
  \bibfield  {author} {\bibinfo {author} {\bibfnamefont {A.}~\bibnamefont {Zenesini}}, \bibinfo {author} {\bibfnamefont {A.}~\bibnamefont {Berti}}, \bibinfo {author} {\bibfnamefont {R.}~\bibnamefont {Cominotti}}, \bibinfo {author} {\bibfnamefont {C.}~\bibnamefont {Rogora}}, \bibinfo {author} {\bibfnamefont {I.~G.}\ \bibnamefont {Moss}}, \bibinfo {author} {\bibfnamefont {T.~P.}\ \bibnamefont {Billam}}, \bibinfo {author} {\bibfnamefont {I.}~\bibnamefont {Carusotto}}, \bibinfo {author} {\bibfnamefont {G.}~\bibnamefont {Lamporesi}}, \bibinfo {author} {\bibfnamefont {A.}~\bibnamefont {Recati}},\ and\ \bibinfo {author} {\bibfnamefont {G.}~\bibnamefont {Ferrari}},\ }\bibfield  {title} {\bibinfo {title} {False vacuum decay via bubble formation in ferromagnetic superfluids},\ }\href {https://doi.org/10.1038/s41567-023-02345-4} {\bibfield  {journal} {\bibinfo  {journal} {Nature Physics}\ }\textbf {\bibinfo {volume} {20}},\ \bibinfo {pages} {558} (\bibinfo {year} {2024})}\BibitemShut {NoStop}%
\bibitem [{\citenamefont {Gleiser}\ \emph {et~al.}(1993)\citenamefont {Gleiser}, \citenamefont {Marques},\ and\ \citenamefont {Ramos}}]{Ramos1993}%
  \BibitemOpen
  \bibfield  {author} {\bibinfo {author} {\bibfnamefont {M.}~\bibnamefont {Gleiser}}, \bibinfo {author} {\bibfnamefont {G.~C.}\ \bibnamefont {Marques}},\ and\ \bibinfo {author} {\bibfnamefont {R.~O.}\ \bibnamefont {Ramos}},\ }\bibfield  {title} {\bibinfo {title} {Evaluation of thermal corrections to false vacuum decay rates},\ }\href {https://doi.org/10.1103/PhysRevD.48.1571} {\bibfield  {journal} {\bibinfo  {journal} {Phys. Rev. D}\ }\textbf {\bibinfo {volume} {48}},\ \bibinfo {pages} {1571} (\bibinfo {year} {1993})}\BibitemShut {NoStop}%
\bibitem [{\citenamefont {Gutierrez~Abed}\ and\ \citenamefont {Moss}(2023)}]{Gutierrez2023}%
  \BibitemOpen
  \bibfield  {author} {\bibinfo {author} {\bibfnamefont {M.}~\bibnamefont {Gutierrez~Abed}}\ and\ \bibinfo {author} {\bibfnamefont {I.~G.}\ \bibnamefont {Moss}},\ }\bibfield  {title} {\bibinfo {title} {Bubble nucleation at zero and nonzero temperatures},\ }\href {https://doi.org/10.1103/PhysRevD.107.076027} {\bibfield  {journal} {\bibinfo  {journal} {Phys. Rev. D}\ }\textbf {\bibinfo {volume} {107}},\ \bibinfo {pages} {076027} (\bibinfo {year} {2023})}\BibitemShut {NoStop}%
\bibitem [{\citenamefont {Baroni}\ \emph {et~al.}(2024)\citenamefont {Baroni}, \citenamefont {Lamporesi},\ and\ \citenamefont {Zaccanti}}]{Baroni2024}%
  \BibitemOpen
  \bibfield  {author} {\bibinfo {author} {\bibfnamefont {C.}~\bibnamefont {Baroni}}, \bibinfo {author} {\bibfnamefont {G.}~\bibnamefont {Lamporesi}},\ and\ \bibinfo {author} {\bibfnamefont {M.}~\bibnamefont {Zaccanti}},\ }\bibfield  {title} {\bibinfo {title} {Quantum mixtures of ultracold gases of neutral atoms},\ }\href {https://doi.org/10.1038/s42254-024-00773-6} {\bibfield  {journal} {\bibinfo  {journal} {Nature Reviews Physics}\ }\textbf {\bibinfo {volume} {6}},\ \bibinfo {pages} {736} (\bibinfo {year} {2024})}\BibitemShut {NoStop}%
\bibitem [{\citenamefont {Cominotti}\ \emph {et~al.}(2023)\citenamefont {Cominotti}, \citenamefont {Berti}, \citenamefont {Dulin}, \citenamefont {Rogora}, \citenamefont {Lamporesi}, \citenamefont {Carusotto}, \citenamefont {Recati}, \citenamefont {Zenesini},\ and\ \citenamefont {Ferrari}}]{Cominotti2023}%
  \BibitemOpen
  \bibfield  {author} {\bibinfo {author} {\bibfnamefont {R.}~\bibnamefont {Cominotti}}, \bibinfo {author} {\bibfnamefont {A.}~\bibnamefont {Berti}}, \bibinfo {author} {\bibfnamefont {C.}~\bibnamefont {Dulin}}, \bibinfo {author} {\bibfnamefont {C.}~\bibnamefont {Rogora}}, \bibinfo {author} {\bibfnamefont {G.}~\bibnamefont {Lamporesi}}, \bibinfo {author} {\bibfnamefont {I.}~\bibnamefont {Carusotto}}, \bibinfo {author} {\bibfnamefont {A.}~\bibnamefont {Recati}}, \bibinfo {author} {\bibfnamefont {A.}~\bibnamefont {Zenesini}},\ and\ \bibinfo {author} {\bibfnamefont {G.}~\bibnamefont {Ferrari}},\ }\bibfield  {title} {\bibinfo {title} {Ferromagnetism in an extended coherently coupled atomic superfluid},\ }\href {https://doi.org/10.1103/PhysRevX.13.021037} {\bibfield  {journal} {\bibinfo  {journal} {Phys. Rev. X}\ }\textbf {\bibinfo {volume} {13}},\ \bibinfo {pages} {021037} (\bibinfo {year} {2023})}\BibitemShut {NoStop}%
\bibitem [{\citenamefont {Recati}\ and\ \citenamefont {Stringari}(2022)}]{Recati2022}%
  \BibitemOpen
  \bibfield  {author} {\bibinfo {author} {\bibfnamefont {A.}~\bibnamefont {Recati}}\ and\ \bibinfo {author} {\bibfnamefont {S.}~\bibnamefont {Stringari}},\ }\bibfield  {title} {\bibinfo {title} {Coherently coupled mixtures of ultracold atomic gases},\ }\href {https://doi.org/10.1146/annurev-conmatphys-031820-121316} {\bibfield  {journal} {\bibinfo  {journal} {Annu. Rev. Condens. Matter Phys.}\ }\textbf {\bibinfo {volume} {13}},\ \bibinfo {pages} {407} (\bibinfo {year} {2022})}\BibitemShut {NoStop}%
\bibitem [{Sup()}]{SupMat}%
  \BibitemOpen
  \href@noop {} {}\bibinfo {howpublished} {{See supplemental material at [URL] for additional details on the sample preparation and data analysis}}\BibitemShut {NoStop}%
\bibitem [{\citenamefont {Abad}\ and\ \citenamefont {Recati}(2013)}]{Abad2013}%
  \BibitemOpen
  \bibfield  {author} {\bibinfo {author} {\bibfnamefont {M.}~\bibnamefont {Abad}}\ and\ \bibinfo {author} {\bibfnamefont {A.}~\bibnamefont {Recati}},\ }\bibfield  {title} {\bibinfo {title} {A study of coherently coupled two-component {B}ose-{E}instein condensates},\ }\href {https://doi.org/10.1140/epjd/e2013-40053-2} {\bibfield  {journal} {\bibinfo  {journal} {The European Physical Journal D}\ }\textbf {\bibinfo {volume} {67}},\ \bibinfo {pages} {148} (\bibinfo {year} {2013})}\BibitemShut {NoStop}%
\bibitem [{\citenamefont {Farolfi}\ \emph {et~al.}(2021)\citenamefont {Farolfi}, \citenamefont {Zenesini}, \citenamefont {Cominotti}, \citenamefont {Trypogeorgos}, \citenamefont {Recati}, \citenamefont {Lamporesi},\ and\ \citenamefont {Ferrari}}]{Farolfi2021}%
  \BibitemOpen
  \bibfield  {author} {\bibinfo {author} {\bibfnamefont {A.}~\bibnamefont {Farolfi}}, \bibinfo {author} {\bibfnamefont {A.}~\bibnamefont {Zenesini}}, \bibinfo {author} {\bibfnamefont {R.}~\bibnamefont {Cominotti}}, \bibinfo {author} {\bibfnamefont {D.}~\bibnamefont {Trypogeorgos}}, \bibinfo {author} {\bibfnamefont {A.}~\bibnamefont {Recati}}, \bibinfo {author} {\bibfnamefont {G.}~\bibnamefont {Lamporesi}},\ and\ \bibinfo {author} {\bibfnamefont {G.}~\bibnamefont {Ferrari}},\ }\bibfield  {title} {\bibinfo {title} {{Manipulation of an elongated internal Josephson junction of bosonic atoms}},\ }\href {https://doi.org/10.1103/PhysRevA.104.023326} {\bibfield  {journal} {\bibinfo  {journal} {Phys. Rev. A}\ }\textbf {\bibinfo {volume} {104}},\ \bibinfo {pages} {023326} (\bibinfo {year} {2021})}\BibitemShut {NoStop}%
\bibitem [{\citenamefont {Lagnese}\ \emph {et~al.}(2021)\citenamefont {Lagnese}, \citenamefont {Surace}, \citenamefont {Kormos},\ and\ \citenamefont {Calabrese}}]{Lagnese2021}%
  \BibitemOpen
  \bibfield  {author} {\bibinfo {author} {\bibfnamefont {G.}~\bibnamefont {Lagnese}}, \bibinfo {author} {\bibfnamefont {F.~M.}\ \bibnamefont {Surace}}, \bibinfo {author} {\bibfnamefont {M.}~\bibnamefont {Kormos}},\ and\ \bibinfo {author} {\bibfnamefont {P.}~\bibnamefont {Calabrese}},\ }\bibfield  {title} {\bibinfo {title} {False vacuum decay in quantum spin chains},\ }\href {https://doi.org/10.1103/PhysRevB.104.L201106} {\bibfield  {journal} {\bibinfo  {journal} {Phys. Rev. B}\ }\textbf {\bibinfo {volume} {104}},\ \bibinfo {pages} {L201106} (\bibinfo {year} {2021})}\BibitemShut {NoStop}%
\bibitem [{\citenamefont {P\^{i}rvu}\ \emph {et~al.}(2024)\citenamefont {P\^{i}rvu}, \citenamefont {Shkerin},\ and\ \citenamefont {Sibiryakov}}]{Pirvu2024}%
  \BibitemOpen
  \bibfield  {author} {\bibinfo {author} {\bibfnamefont {D.}~\bibnamefont {P\^{i}rvu}}, \bibinfo {author} {\bibfnamefont {A.}~\bibnamefont {Shkerin}},\ and\ \bibinfo {author} {\bibfnamefont {S.}~\bibnamefont {Sibiryakov}},\ }\href@noop {} {\bibinfo {title} {Thermal false vacuum decay is not what it seems}} (\bibinfo {year} {2024}),\ \Eprint {https://arxiv.org/abs/2407.06263} {arXiv:2407.06263 [hep-th]} \BibitemShut {NoStop}%
\bibitem [{\citenamefont {Maki}\ \emph {et~al.}(2023)\citenamefont {Maki}, \citenamefont {Berti}, \citenamefont {Carusotto},\ and\ \citenamefont {Biella}}]{Maki2023}%
  \BibitemOpen
  \bibfield  {author} {\bibinfo {author} {\bibfnamefont {J.~A.}\ \bibnamefont {Maki}}, \bibinfo {author} {\bibfnamefont {A.}~\bibnamefont {Berti}}, \bibinfo {author} {\bibfnamefont {I.}~\bibnamefont {Carusotto}},\ and\ \bibinfo {author} {\bibfnamefont {A.}~\bibnamefont {Biella}},\ }\bibfield  {title} {\bibinfo {title} {{Monte Carlo matrix-product-state approach to the false vacuum decay in the monitored quantum Ising chain}},\ }\href {https://doi.org/10.21468/SciPostPhys.15.4.152} {\bibfield  {journal} {\bibinfo  {journal} {SciPost Phys.}\ }\textbf {\bibinfo {volume} {15}},\ \bibinfo {pages} {152} (\bibinfo {year} {2023})}\BibitemShut {NoStop}%
\bibitem [{Dat()}]{Data}%
  \BibitemOpen
  \href@noop {} {}\bibinfo {howpublished} {{https://zenodo.org/records/17279521}}\BibitemShut {NoStop}%
\end{thebibliography}

%

\renewcommand{\thefigure}{S\arabic{figure}}

\setcounter{figure}{0}
\section*{Supplemental Material}

\section{Preparation of metastable ferromagnetic systems}

\begin{figure}[b!]
    \centering
    \includegraphics[width=1.05\columnwidth]{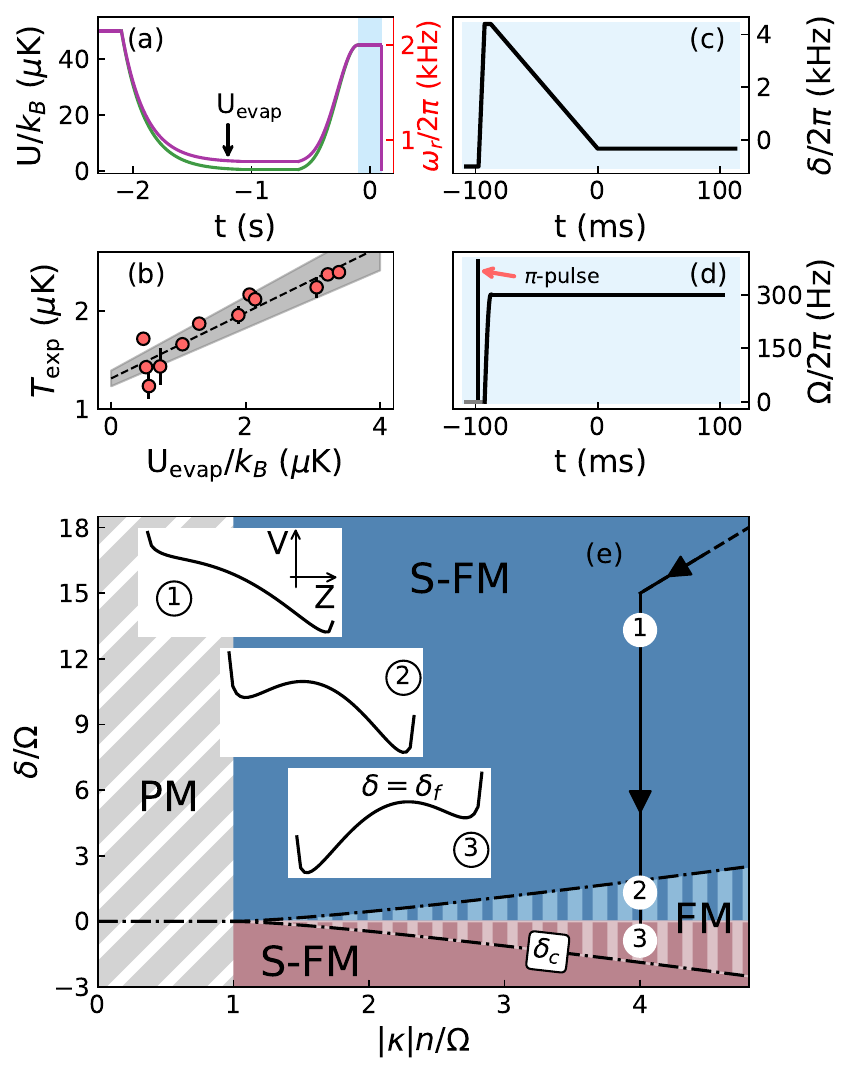}
    \caption{Experimental sequence. (a) Variation in time of the optical potential $U$. The temperature of the sample can be tuned by varying the trapping potential $U_{\mathrm{evap}}$ at the bottom of the evaporation ramp. The trap is later recompressed in 500 ms to a final radial trapping frequency of 2 kHz. Panel (b) shows that the temperature measured after the recompression ramp varies linearly with the value of $U_{\mathrm{evap}}$. The time sequence of detuning and coupling strength is shown in panel (c) and (d), respectively. For comparison, $|\kappa|n \simeq $ 1 kHz. (e) Phase diagram of the ferromagnetic superfluid mixture showing the paramagnetic (PM), ferromagnetic (FM), and saturated ferromagnetic (S-FM) regions as a function of the detuning and spin interactions in units of the coupling strength. The color code is the same used in Fig.\,2 of the main text (blue for $\up$ and red for $\down$). The black line illustrates the path followed by the mixture during the preparation process. The three insets show the potential energy profile in different regions. }
    \label{fig:figS1}
\end{figure}

A coherently-coupled sodium mixture in the two spin states $|F,m_F\rangle=\,$\mtwo $=\,\up$ and \mone=\,$\down$ is prepared following the protocol illustrated in \aref{fig:figS1}.
The protocol consists in two steps: a first part dedicated to the preparation of a partially-Bose-condensed cloud, shown in \aref{fig:figS1}(a), followed by a spin-manipulation scheme to create a metastable ferromagnet, \aref{fig:figS1}(c) and \aref{fig:figS1}(d).

We start with a sample polarized in \down at $\SI{10}{\mu K}$ and evaporatively cool it down below the condensation critical temperature in a single-beam optical dipole trap, by exponentially reducing the optical potential $U$ down to a variable minimum value $U_{\mathrm{evap}}$ [\aref{fig:figS1}(a)]. After a thermalization time of 500 ms, $U$ is increased to a final fixed value of about $ k_B\times\SI{45}{\mu K}$, corresponding to an elongated harmonic trap with axial and radial trapping frequencies  $\{\omega_x,\omega_r\}/2\pi=\{15(1), 2020(20)\}$\,Hz.
In this way, despite a small heating  during the recompression, at the end of the procedure the atomic sample is always confined in the same optical trap, but can have different temperatures $T_\mathrm{exp}$, tuned by the choice of $U_{\mathrm{evap}}$ [\aref{fig:figS1}(b)], as explained in detail later in the text.

After recompression, a resonant and fast (20-$\mu$s-long) $\pi$-pulse with $\Omega = 2\pi \times\SI{25}{kHz}$  transfers all atoms to the $\up$ state.
The detuning $\delta$, which includes the frequency mismatch between the two-level system and the microwave radiation, as well as the density-dependent collisional shift of $\sim 2 \pi \times  \SI{1}{kHz}$, is ramped up to a high positive value $\sim 2 \pi \times  \SI{4.4}{kHz}$ [see \aref{fig:figS1}(c)]. In this way, the potential energy $V(Z)$ in Eq.\,2 of the main text has just one minimum, at any value of the coupling strength $\Omega$. 
At this stage, we switch on the coupling and ramp its strength $\Omega$ up to $\sim 2\pi \times \SI{300}{Hz}$ in 5\,ms, after which we ramp down $\delta$ to a variable value $\delta_f$ at a speed of 54\,Hz/ms  [\aref{fig:figS1}(c),(d)]. We then wait a variable time $t$ before applying the spin-selective absorption imaging.

\section{Image acquisition and analysis}
\paragraph{Image acquisition}
At time $t$ after the end of the ramp on $\delta$, we switch the coupling off and release the mixture from the trap. After $\SI{1}{ms}$ of time of flight, we acquire a first image of atoms in the $\up$ state, directly on the closed optical transition $|2,-2\rangle\rightarrow |3,-3\rangle$. After an additional ms, we transfer the atoms in the $\down$ state to the $\up$ state with a 20-$\mu$s-long microwave $\pi$-pulse, and acquire the second image on the closed transition to detect the density distribution of the $\down$ state. 

\begin{figure}[t]
    \centering
    \includegraphics[width=1.1\columnwidth]{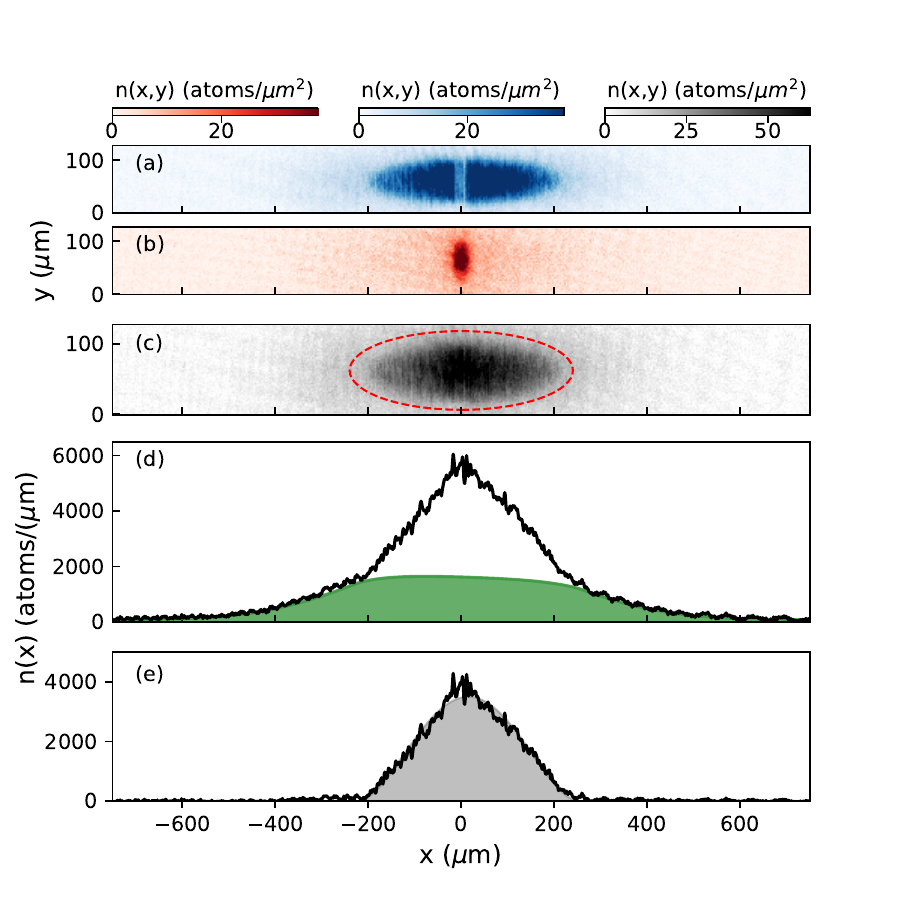}
    \caption{Imaging processing sequence for BEC parameter extraction. Two-dimensional ODs of the \up (a) and \down (b) atoms, and total density $n(x,y)$ (c). The red dashed line in panel (c) represents the boundary between the BEC region (inside, black) and thermal region (outside, dark gray) identified after the 2D bimodal fit. (d) Radially integrated total density $n(x)$ from the 2D density reported in panel (c). The thermal part, integrated radially, is highlighted in green. (e) Total density after thermal part removal. The grey area is the Thomas-Fermi fit used to extract $\kappa n$.}
    \label{fig:FigS2}
\end{figure}

Due to the different times of flight and the strong radial confinement, the radial distributions of the two spin states are not the same, narrower for the gas originally in $\up$ and broader for the gas in $\down$.
We correct this effect in post-processing by stretching the radial direction of the optical density (OD) image of $\up$ to match the radial dimension of the other one, see \aref{fig:FigS2}(a) and (b). The expansion in the longitudinal direction can instead be neglected due to the time of flight being much shorter than the characteristic trap period of about $\SI{66}{ms}$.

\paragraph{Image analysis}
Since the atomic distribution contributing to the OD contains information on both the condensed and thermal fractions, it is crucial to separate them. This is particularly important for the calculation of the condensate magnetization $Z$, because in general the spin dynamics of the thermal gas will be different from the one of the ferromagnetic superfluid. In the following, we present the step-by-step procedure applied to isolate the condensed fraction.

We first identify the region occupied by the condensate, delimited by the Thomas-Fermi radii, by fitting the total OD with a bimodal function containing the following components:
\begin{eqnarray}
n_{B}(x,y) = & \label{eq:BEC2D}  \\
n_{0, B} &\max\left\{\left[1 -\left(\frac{x-x_{0,B}}{R_x}\right)^2 -\left(\frac{y-y_{0, B}}{R_y}\right)^2\right]^{3/2},0 \right\} \nonumber
\end{eqnarray}

\begin{eqnarray}
n_{th}(x,y) =& \label{eq:thermal}\\
n_{0, th}\,& g_2 \left\{ 
\exp{\left[ -\frac12\left(\left(\frac{x-x_{0, th}}{\sigma_x}\right)^2 + \left(\frac{y-y_{0, th}}{\sigma_y}\right)^2\right)\right]}\right\}. \nonumber
\end{eqnarray}

Here $n_B(x,y)$ and $n_{th}(x,y)$ represent the 2D distributions for the condensate and thermal components of the system. The two distributions are parametrized by their amplitudes $n_{0, B/th}$, their centers $x_{0, B/th}$ and $y_{0, B/th}$, the TF radii $R_x$ and $R_y$, and the thermal widths $\sigma_x$ and $\sigma_y$. In addition a global offset is used in order to take into account possible imaging imperfections.
In \aref{eq:thermal} the function $g_2$ is the polylogarithm function of index 2. Notice that in general the \textit{BEC center} does not coincide with the center of the thermal cloud.

Once the condensate region in the total density is identified,
we separately treat the thermal component inside and outside the boundaries of the condensate region, see red dashed line in \aref{fig:FigS2}(c).
For each spin state, the thermal component outside the condensate is identified by fitting $n_{th}(x,y)$ to the OD in the region outside the ellipse delimited by $R_x$ and $R_y$.
Inside the ellipsoidal region occupied by the BEC, the thermal components are depleted, and their integration along the line of sight provides, to a good approximation, a flat contribution \cite{Cominotti2023}.
Given the different locations of the centers of the thermal and condensed parts, we consider the inner thermal component as a plane, tilted along the $x$-axis, which interpolates between the values of the Bose function $n_{th}$ at $(x, y) = (x_{0, B} \pm R_x,0)$. 

The reconstructed thermal profiles in the outer and inner regions are then separately subtracted from the ODs of the two states, allowing us to obtain the BEC densities $n_{\downarrow}$ and $n_{\uparrow}$. From the latter, we determine the density $n(x)=(n_{\uparrow}+n_{\downarrow})$ and the magnetization $Z(x)=(n_{\uparrow}-n_{\downarrow})/(n_{\uparrow}+n_{\downarrow})$, after integration of $n_{\downarrow}$ and $n_{\uparrow}$ along the radial direction. 
Figure\,\ref{fig:FigS2}(d) reports the integrated total density with the integrated thermal profile highlighted in green. Subtracting the thermal contribution from the total integrated density, we get the total condensed distribution [\aref{fig:FigS2}(e)].

We fit the integrated density with an integrated 1D Thomas-Fermi profile,  $n(x) = n_{1D} (1-(x-x_0)^2/R_x^2)^2$, where $n_{1D}$ is the the 1D density, and use the resulting fit parameters to calculate the spin interaction energy \cite{Cominotti2023}

\begin{equation}
    \kappa n(x) = \frac23\left(\frac{g_{\downarrow\downarrow} + g_{\uparrow\uparrow}- 2 g_{\uparrow\downarrow}}{2\hbar} \right)n_0^{3D}\left(1-\frac{x^2}{R_x^2}\right),
    \label{eq:eqkn}
\end{equation}

with $g_{ij} = 4\pi\hbar^2 a_{ij}/m$ the interaction constants between two atoms in the state $\ket{i}$ and $\ket{j}$, respectively, and $n_0^{3D} = 15 N/(8\pi R_xR_r^2)$ the 3D peak density of a sample of $N$ atoms. For the mixture used in the experiment, 
$a_{\uparrow\uparrow} = 64.3\, a_0$, $a_{\downarrow\downarrow} = 54.5\, a_0$, $a_{\uparrow\downarrow} = 64.3\, a_0$. The exact value of $N$ can be extracted from the fit values using the expression

\begin{equation}
   n_{1D} = \frac{15}{16}\frac{N}{R_x},
    \label{eqs:n1d}
\end{equation}

which originates from the integration of the density of a 3D condensate along two orthogonal lines of sight.
We independently measured the two trapping frequencies and extracted the trap aspect ratio as  $\omega_x/\omega_r = 0.0075$. We use this information to evaluate  $R_r =  R_x \omega_x/\omega_r$, which is not directly measurable within the optical resolution of our imaging system. 

Note that, as detailed in Ref.~\cite{Cominotti2023, Farolfi2021}, \aref{eq:eqkn} indicates that $\kappa n$ is proportional to the 3D peak density of the condensate, but is distributed as a 1D Thomas-Fermi profile. This originates from a dimensionality reduction, under the assumption that the radial dynamics of the magnetization $Z$ is suppressed. 
Moreover, the value of $\kappa n$ is used to calculate the spin healing length $\xi_s$ as

\begin{equation}
      \xi_s  = {\sqrt\frac{\hbar}{2m |\kappa| n}}.
    \label{eqs:xi}  
\end{equation}

Uncertainties on $|\kappa|n$, $n_{1D}$, and  $\xi_s$ are determined \textit{a posteriori} from the standard deviations over the different experimental realizations at fixed $\delta_f$ and $\Omega$ where no bubble is detected, see next section.

\section{Bubble Identification and Determination of {\boldmath $\tau$ }}
For each experimental realization, we analyze the magnetization profile and establish the presence of spin bubbles at a given time $t$ when $Z(x)$ contains a region of at least $\SI{10}{\mu m}$ where $Z$ is smaller than a set threshold. We set this threshold to $Z=0.2$, as in Ref.~\cite{Zenesini2024}. 

As described in the main text, the characteristic decay rate is computed after the evaluation of the FV survival probability
\begin{equation}
    F_t=\frac{\langle Z(x) \rangle_t-Z_{TV}}{\langle Z(x)\rangle_0-Z_{TV}},
\end{equation}
where $\langle \cdot \rangle$ is the spatial average in the central $\SI{60}{\mu m}$ over several realizations.
Due to decoherence and experimental noise, the exact value of $Z_{TV}$ does not necessarily correspond to the theoretical one. Nevertheless, under the assumption that the system is well initialized in the FV state, $\langle Z\rangle_0 = Z_{FV}$, and that the false and true vacuum correspond to opposite magnetization, the approximation $Z_{TV} = - \langle Z\rangle_0$ allows to write
\begin{equation}
    F_t=\frac12 \left(1 +\frac{\langle Z(x) \rangle_t}{\langle Z(x)\rangle_0}\right),
\end{equation}
which is more direct to compute from the experimental data.
We start averaging the magnetization $Z(x)$, for each set of ($\delta_f-\delta_c$) and temperature $T_\mathrm{exp}$, by grouping the experimental realizations in 10 equally spaced intervals, according to the acquisition time $t$. 
For each pair of values of $\delta_f$ and $T_\mathrm{exp}$,  $\langle Z \rangle_0 $ is obtained from the average magnetization over all experimental realizations without bubbles. 

\section{Determination of {\boldmath  ($\delta_f-\delta_c$) } }

When the whole ferromagnetic mixture is prepared in the metastable configuration, it decays through the generation of bubbles with a characteristic time, which is longer the larger is the difference between $\delta_f$ and the critical detuning $\delta_c$.

The determination of ($\delta_f-\delta_c$) relies on the exact identification of the condition $\delta=0$ and on the knowledge about how $\delta_c$ varies with $|\kappa| n$, as can be seen in \aref{fig:figS1}(e). 
This is performed in two steps:\\
\noindent 1) For a (non interacting) thermal cloud, the resonant condition $\delta=0$ is determined with very high precision thanks to Rabi spectroscopy before and after each FVD acquisition series. 
Relative to such resonant condition, we set different values $\delta_f$ for each experimental run to explore changes in the bubble formation dynamics.
The uncertainty on $\delta_f$ comes only from the Rabi spectroscopy accuracy $\sim 2 \pi \times$5 Hz and from small field drifts during the acquisition time $\sim 2 \pi \times$5 Hz.

\noindent 2) In an ideal 1D system the critical detuning $\delta_c$ has an analytical dependence on the experimental parameters $\kappa n$ and $\Omega$. In Ref.~\cite{Cominotti2023} we found that this is not applicable to our quasi-1D condensate. In the present work, as in Ref.~\cite{Zenesini2024}, we rely on the experimental observation of the threshold value of $\delta_f$ at which the lifetime of the false vacuum is compatible with zero, marking the disappearance of the second minimum of $V(Z)$. The protocol consists in measuring $Z(x)$ at $t=0$ with the same sequence used for bubble detection, and determining at which detuning the state \down starts to be observed. We average tens of these scans and, in the investigated range of $\delta_f$, we find the following linear dependence
\begin{equation}
    \delta_c=0.35~|\kappa| n+2\pi\times 280~\mathrm{Hz},
\end{equation}
valid for $\Omega = 2\pi \times \SI{300}{Hz}$.
This allows us to obtain $\delta_c$ for each FVD scan from the value of $|\kappa| n$ determined from the ODs, as explained in the  previous section.  We also observed no dependence of $\delta_c$ on $T_\mathrm{exp}$. 

We estimate the uncertainty on $(\delta_f-\delta_c)$ to be on the order of $2\pi \times 20$~Hz. Any systematics in the determination of $\delta_c$ can introduce a common shift equal for all the extracted values of $\varepsilon_c$ in the main text. 
This can affect the determination of the value of $a$ by a scaling factor, but does not affect the parameter $b$.

\begin{figure}[t!]
    \centering
    \includegraphics[width=0.9\columnwidth]{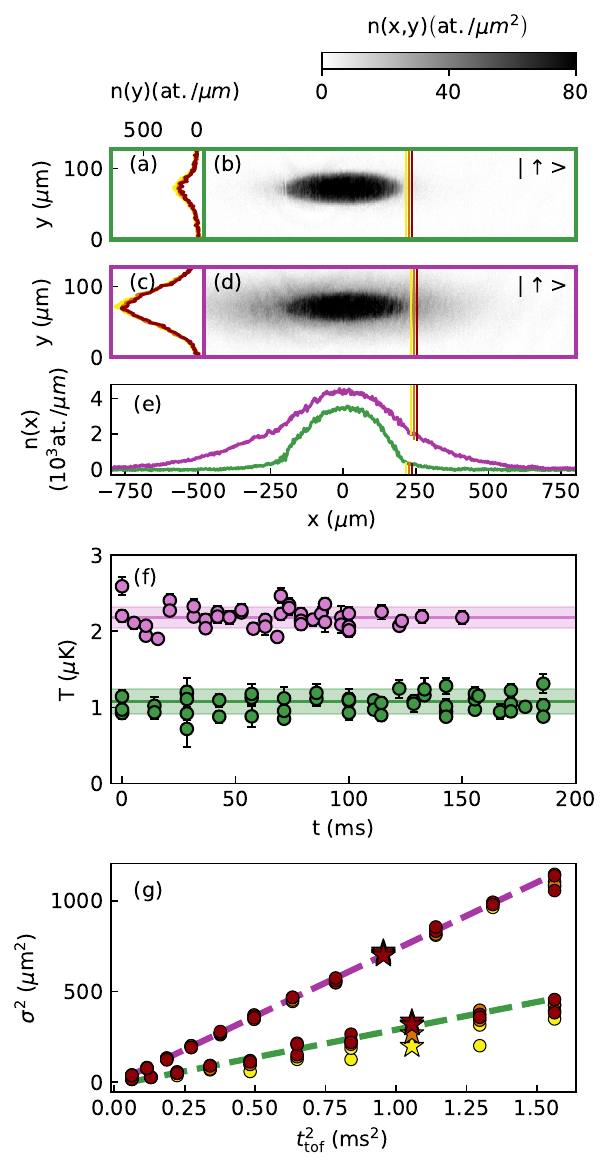}
    \caption{Determination of the sample temperature. Pictures of a cold (b, green) and a hot (c, purple) atomic cloud in the starting spin state and their radially integrated density (e). For each experimental image, we determine the vertical distribution of atomic density [(a) and (c)] for three different axial positions outside the Thomas Fermi radius (yellow, orange, and red line). The distributions are fit to a Bose function to obtain three values of temperature, which are averaged to determine $T_\mathrm{exp}$ for each experimental realization. (f) Example of evolution in time of the sample temperature for two different datasets, with sufficiently long $\tau$. For each dataset, the temperature is normalized to its average value $T_\mathrm{exp}$. Values of $T_\mathrm{exp}$ are $\SI{1.1(2)}{\mu K}$, $\SI{2.2(1)}{\mu K}$ for the green and purple points, respectively. To validate the procedure, we verify that the temperature extracted in this way agrees with the one obtained with the standard time-of-flight procedure (g). Different points come from different repetitions, and stars mark the values of $\sigma^2$ from pictures shown in (b) and (d). }
    \label{fig:FigS3}
\end{figure}

\begin{figure*}[t!]
    \centering
    \includegraphics[width=\textwidth]{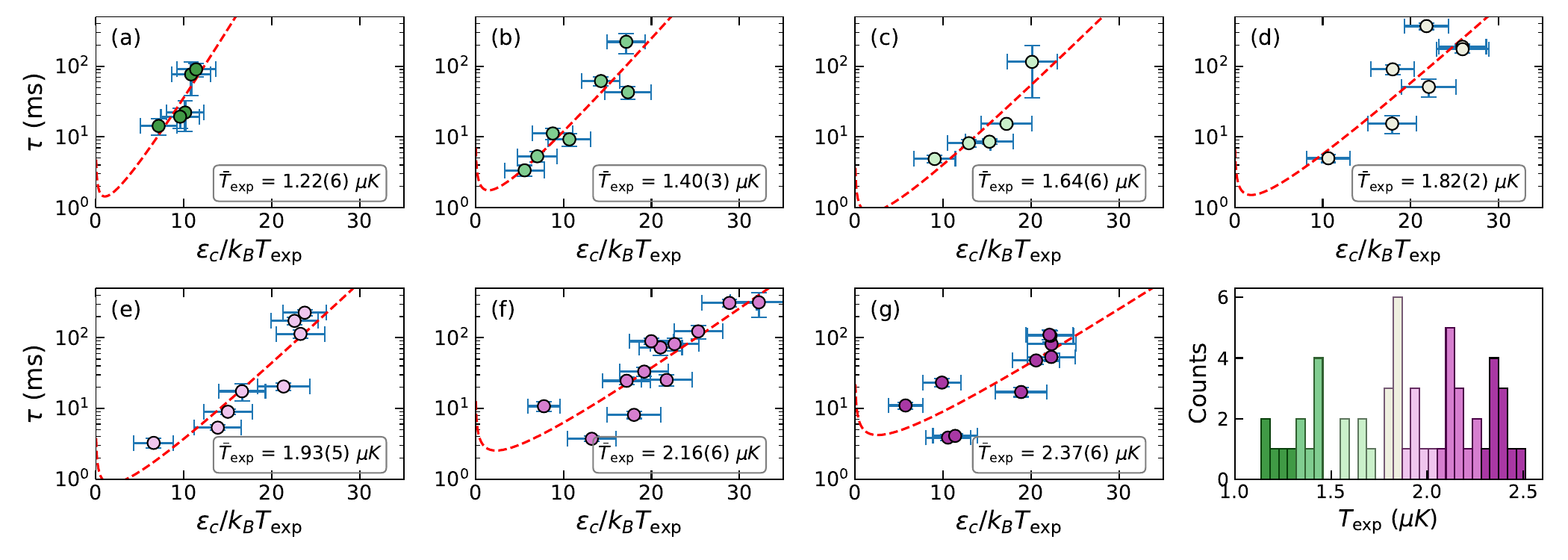}
    \caption{(a-g) Measurement of the FV decay time $\tau$ as a function of the rescaled instanton energy in units of $k_BT_\mathrm{exp}$ for different temperatures (reported in each box). The dashed lines are the fits to the data according to the instanton model. (h) Data distribution as a function of the measured temperature.}
    \label{fig:FigS4}
\end{figure*}

\section{Temperature determination}

For each dataset at fixed ($\delta_f-\delta_c$), we extract the temperature $T_\mathrm{exp}$ analyzing all images taken at waiting times less than $\SI{200}{ms}$, which do not contain any bubble. We directly extract the temperature from the short time-of-flight images discussed in the second section. 
For each sample, we consider the total OD along the radial direction $y$ at three different axial positions outside the condensate $x_T/R_x = {1.05, 1.1, 1.15}$ (respectively yellow, orange, and red in \aref{fig:FigS3}), fit them to an integrated 1D Bose function and extract the width of the distribution $\sigma(x_T)$. Since the in-trap width of the thermal distribution relates to the radial curvature of the optical trap, and the latter varies along the axial direction $x$, we evaluate $\omega_r(x_T)$ by means of Gaussian beam propagation. The temperature $T_{\mathrm{exp},i}$, associated to the \textit{i}-th experimental realization, is calculated as the average of the three temperatures evaluated at each position
\begin{equation} 
    T_{\mathrm{exp}, i} = \frac{1}{3}\sum_{x_T} \frac{m\sigma^2(x_T)}{k_B} \frac{\omega_r^2(x_T)}{1+\omega_r^2(x_T)t_\mathrm{tof}^2},  
\end{equation}
where $t_\mathrm{tof}$ is the time-of-flight between releasing the atoms from the trap and the imaging acquisition. 
With this method, we do not observe any significant variation in the temperature at different waiting times $t$, as illustrated in \aref{fig:FigS3}(f). Finally, the value of $T_{\mathrm{exp}}$ associated to a single value of $\tau$ is obtained by averaging $T_{\mathrm{exp},i}$, and the corresponding uncertainties are determined from the standard deviation.

We verified that this single-shot thermometry is reliable and consistent with the standard time-of-flight expansion, see \aref{fig:FigS3}(g), but with the great advantage of allowing to extract the temperature without a separated measurement.

\section{Determination of {\boldmath $b$} and linear dependence on {\boldmath $T_\mathrm{exp}$ } }

In \aref{fig:FigS4}, we report the complete set of fits of the instanton theory on the data reported in the main text [see panels (a-g)] for the different temperatures. Data in panels (b) and (f) correspond to those shown in Fig.~3 of the main text. We checked that the linearity of $b^{-1}$ on $T_\mathrm{exp}$ is maintained when the number of clusters changes to 6 or 8, as well as the compatibility of $a$ with a constant value within experimental uncertainty.

\end{document}